\documentclass[pre,reprint,superscriptaddress,amsmath]{revtex4-2}
\usepackage{newtxtext,newtxmath}
\usepackage[colorlinks]{hyperref}
\usepackage{graphicx}

\begin{document}
\title{Denoising Scheme Based on Singular-Value Decomposition for One-Dimensional Spectra and Its Application in Precision Storage-Ring Mass Spectrometry}
\author{X.C.~Chen}
\email{cxc@impcas.ac.cn}
\affiliation{Institute of Modern Physics, Chinese Academy of Sciences, Lanzhou 730000, China}
\author{Yu.~A.~Litvinov}
\affiliation{Institute of Modern Physics, Chinese Academy of Sciences, Lanzhou 730000, China}
\affiliation{GSI Helmholtzzentrum f\"ur Schwerionenforschung GmbH, 64291 Darmstadt, Germany}
\affiliation{Max-Planck-Institut f\"ur Kernphysik, 69117 Heidelberg, Germany}
\author{M.~Wang}
\affiliation{Institute of Modern Physics, Chinese Academy of Sciences, Lanzhou 730000, China}
\author{Q.~Wang}
\affiliation{Institute of Modern Physics, Chinese Academy of Sciences, Lanzhou 730000, China}
\affiliation{University of Chinese Academy of Sciences, Beijing 100049, China}
\author{Y.H.~Zhang}
\affiliation{Institute of Modern Physics, Chinese Academy of Sciences, Lanzhou 730000, China}

\date{\today}
\begin{abstract}
  This work concerns noise reduction for one-dimensional spectra in the case that the signal is corrupted by an additive white noise.
  The proposed method starts with mapping the noisy spectrum to a partial circulant matrix.
  In virtue of singular-value decomposition of the matrix, components belonging to the signal are determined by inspecting the total variations of left singular vectors.
  Afterwards, a smoothed spectrum is reconstructed from the low-rank approximation of the matrix consisting of the signal components only.
  The de-noising effect of the proposed method is shown to be highly competitive among other existing non-parametric methods including moving average, wavelet shrinkage, and total variation.
  Furthermore, its applicable scenarios in precision storage-ring mass spectrometry are demonstrated to be rather diverse and appealing.
\end{abstract}
\maketitle

\section{Introduction}
Noise is ubiquitous in the outcome of any physical experiment, owing to statistical fluctuations and various uncontrollable disturbances affecting the measuring system.
It is superposed over the measured signal, and obscures the inference of the underlying ground truth.
To reduce the noise effect, usually a batch of independent measurements under the same condition are performed for averaging, since the noise variance scales inversely with the averaging number.
However, the accumulated statistics may be insufficient to bring down the noise due to pragmatic constraints.
Or even the measurement is just a one-shot instance.
Under those circumstances, data de-noising in the later analysis can only be resorted to as a mitigation.

It is quite common that measurement results are encoded in one-dimensional spectra, such as the energy spectrum of a $\gamma$-radioactive isotope, the frequency response of a notch filter, or the hourly temperature readings of a thermometer.
Mathematically speaking, it is an ordered sequence of real-valued numbers, where usually the signal is smooth while the noise overlays complex irregularities like kinks and wiggles.
In this sense, smoothing can be used interchangeably with de-noising in the context of one-dimensional data analysis.

Despite the voluminous data smoothing techniques, there are in general two categories, namely parametric and non-parametric.
A parametric method starts with a data model built from \textit{a priori} knowledge on the signal and the noise.
Such knowledge can be a theoretical comprehension of the measuring system, or merely an educated guess.
In contrast, a non-parametric method makes a minimal assumption about the noise and then lets the data speak for themselves.
Thus with less restrictions, non-parametric methods own more versatility and applicability.

As an example in the non-parametric category, moving average probably is the most intuitive method~\cite{hyndman_moving_2011}.
The smoothed sequence is formed by replacing every noisy datum with the (weighted) mean of data in its neighborhood.
This is essentially equivalent to convolution smoothing~\cite{clark_non-parametric_1977}, where the noisy sequence is convolved with a smoothing kernel such as binomial, Gaussian, or exponential~\cite{roberts_control_1959}.
In the same vein, the neighborhood for averaging can be extended to encompass data with similar values, which has inspired two other methods: bilateral filter~\cite{tomasi_bilateral_1998} and non-local means~\cite{buades_non-local_2005}.

When viewed from a different perspective, the convolution smoothing is in fact low-pass filtering in the Fourier-transformed domain~\cite{ullmann_picture_1976}, where the smoothing kernel is the impulse response of the filter.
Normally the signal is band-limited whereas the noise covers the whole domain, hence a low-pass filter can effectively reject most of the noise.
Based on the same Fourier transform, another method named Fourier thresholding takes a different approach to data de-noising~\cite{huang_noise_1974}.
It abandons sinusoidal components whose transform coefficients are below a given threshold, as they more likely belong to the noise.

Like Fourier transform, wavelet transform is another powerful tool in digital signal processing~\cite{daubechies_ten_1992}, which decomposes an input sequence into localized oscillatory components.
The Fourier-based data de-noising methods can likewise be migrated to the wavelet-transformed domain.
As a counterpart of the Fourier thresholding, wavelet shrinkage manipulates wavelet coefficients according to a specific criterion to smooth a noisy sequence~\cite{donoho_ideal_1994,donoho_wavelet_1995}.

Lastly, it is worth noting a special mainstream method in the non-parametric category: total variation regularization~\cite{rudin_nonlinear_1992}, whose underlying framework differs from the aforementioned.
The idea is to minimize the total variation, which is the $\ell_1$-norm of its first derivative, of the smoothed sequence.
Meanwhile, it also tries to preserve data fidelity as the regularizing condition.
Since the $\ell_1$- rather than $\ell_2$-norm is adopted, this method is good at retaining discontinuities in the noisy sequence~\cite{strong_edge-preserving_2003}, but also suffers from a staircase effect in the smoothed one~\cite{caselles_discontinuity_2007}.

On the contrary, curve fitting of a pre-defined signal model by least-squares regression is a typical data de-nosing example in the parametric category.
To apply the fitting, the noisy sequence can be treated as a whole or in partitions, where the latter gives rise to locally weighted regression~\cite{cleveland_robust_1979} and Savitzky-Golay filter~\cite{savitzky_smoothing_1964}.
If additionally a noise model is defined, data de-noising can be achieved in a more sophisticated manner.
Examples are Wiener filter~\cite{wiener_extrapolation_1949} and Kalman filter~\cite{kalman_new_1960}, which are both rooted in statistical inference and estimation.

In the work presented here, we propose a non-parametric one-dimensional data de-noising method based on Singular-Value Decomposition (SVD).
The SVD is a well-developed matrix factorization technique~\cite{stewart_early_1993}, which can be used to realize the idea of principal component analysis~\cite{pearson_liii._1901}.
Its applications are prevalent in digital signal processing, especially in two-dimensional image-related works such as coding, watermarking, and de-noising~\cite{andrews_singular_1976,chang_svd-based_2005,guo_efficient_2016,furnival_denoising_2017}.
With the aid of a bijective map from a sequence to a matrix, the SVD also plays a role in one-dimensional data analysis, such as speech recognition~\cite{hermus_fully_1999}, fetal electrocardiogram extraction~\cite{al-zaben_extraction_2006}, mass spectrometry~\cite{chiron_efficient_2014}, and fault diagnosis in mechanical systems~\cite{cong_short-time_2013,zhao_novel_2017,zhao_selection_2011,jiang_study_2015}.
In particular for the data de-noising, a noisy sequence is to be decomposed by SVD, where a smoothed sequence can be reconstructed from a selected subset of the components.

\section{Method}
Let us consider a noise-corrupted sequence $x$ of length $n$, which consists of the signal $a$ and the noise $d$:
\begin{equation}
  x_i = a_i + d_i,\quad 0\leq i < n.
  \label{eq:add}
\end{equation}
Here, we have made no assumptions about the noise other than being additive, zero-mean, and white.

The proposed data de-noising method mainly includes three steps: (1) Embed the sequence into a matrix. (2) Apply SVD to the matrix and determine signal components. (3) Reconstruct a smoothed sequence with the selected components.
In the following, each step will be elaborated.

\subsection{Matrix selection}
In literature there are mainly three approaches to embed the sequence $x$ into a matrix.
Probably the simplest one is to evenly partition $x$ into segments, which are then stacked to form a matrix~\cite{schanze_compression_2018,al-zaben_extraction_2006,cong_short-time_2013}.
This approach is mostly suitable for a quasi-periodic sequence, when each segment roughly contains one or multiple periods.
Another approach is to build a time-frequency representation of $x$~\cite{hassanpour_timefrequency_2008}, which can be computationally expensive.
As a matter of fact, the frequently adopted approach is to construct a Hankel matrix~\cite{zhao_novel_2017,chiron_efficient_2014,gong_improved_2017,hermus_fully_1999,shin_iterative_1999,zhao_selection_2011,jiang_study_2015}, which balances algorithmic performance and computational cost.

Specifically, the constructed Hankel matrix $H$ of size $m\times(n-m+1)$ reads
\begin{equation}
  H_{ij} = x_{i+j}, \quad 0\leq i<m;\, 0\leq j\leq n-m.
  \label{eq:hankel}
\end{equation}
Apparently, the choice of $m$ will somehow influence the smoothing effect.
It is shown that $m$ should best be $n/2$ or $(n+1)/2$, depending on the parity of $n$~\cite{zhao_selection_2011}.
That is to say, $H$ should optimally be (almost) square.

However, a drawback of the mapping in Eq.~(\ref{eq:hankel}), which is often overlooked, is that the elements of $x$ are not put on an equal footing, i.e., the central elements tend to occur more times in $H$ than the lateral ones.
Numerical experiments have proved that by construction of $H$ the signal details close to the sequence ends can badly be recovered after de-noising.

\begin{figure}
  \includegraphics[width=.9\columnwidth]{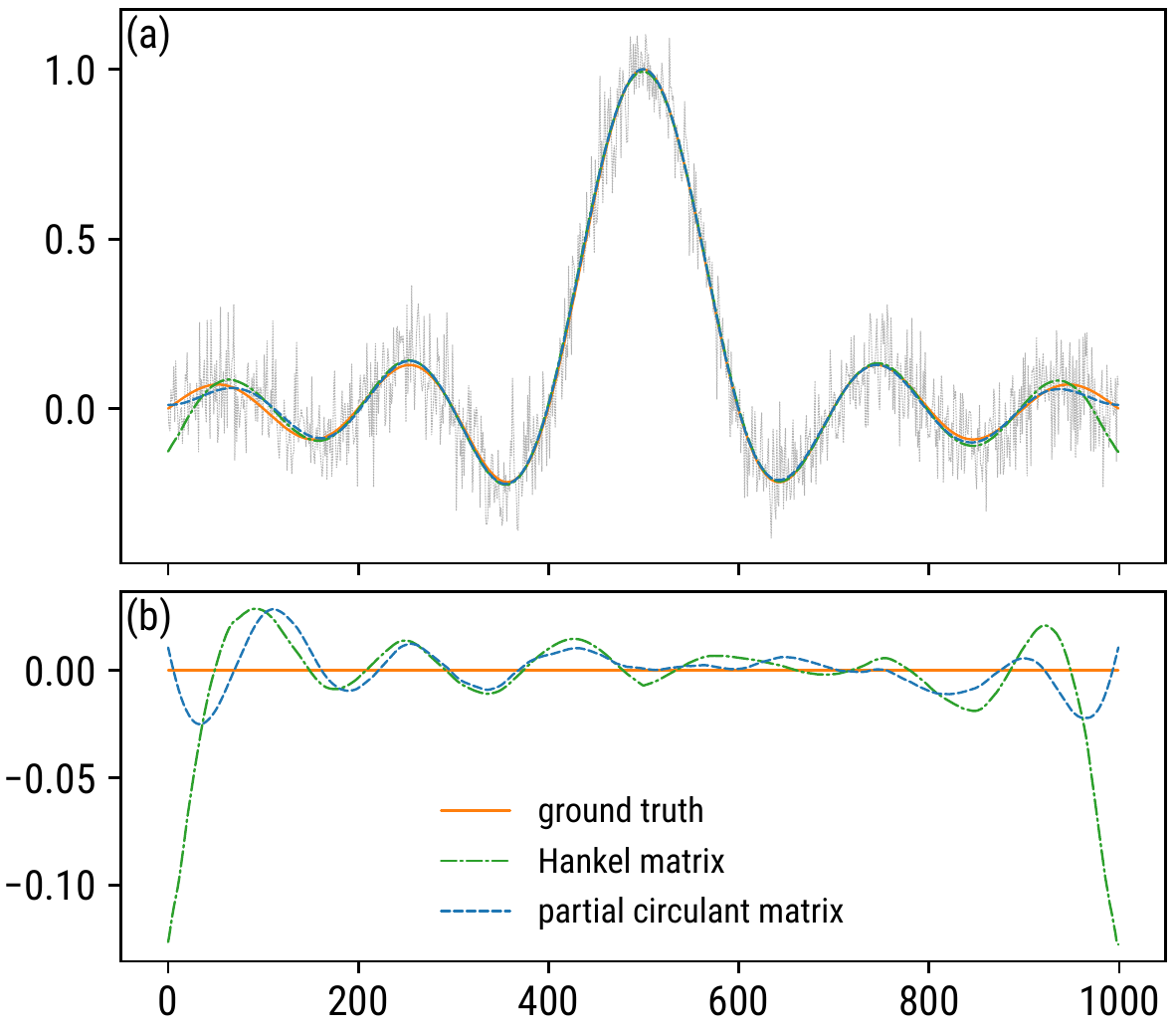}
  \caption{
    Illustration of different de-noising effects with two matrix-embedding approaches.
    (a) The green line shows the de-noised result with a Hankel matrix, while the blue line corresponds to a partial circulant matrix.
    The orange line is the ground truth.
    (b) Deviations of the two results with respect to the ground truth.
    \label{fig:drawback}
  }
\end{figure}

Fig.~\ref{fig:drawback} illustrates such a phenomenon with a synthetic sequence, which is the superposition of a sinc signal and a normal noise.
The sequence of length $n=1000$ is embedded into $H$ with a row number $m=500$.
It is evident that the de-noised sequence (see green line in Fig.~\ref{fig:drawback}) strongly deviates from the ground truth (see orange line in Fig.~\ref{fig:drawback}) at both ends, where the largest deviation amounts to $10.5\%$.
What is worse, the deviation is so large that the de-noised sequence is notably off the noisy one, which indicates the inability to respect lateral information in the case of de-noising with $H$.

To overcome this limitation, we thus propose to construct a partial circulant matrix $X$ of size $m\times n$ with $m \leq n$:
\begin{equation}
  X_{ij} = x_{(i+j)\bmod{n}}, \quad 0\leq i<m;\, 0\leq j<n.
  \label{eq:pcm}
\end{equation}
Every row of $X$ is a cyclic left-shift of the above one by an element, such that every $x_i$ occurs exactly $m$ times.
By virtue of $X$ with the same row number $m=500$, the de-noised sequence, which is shown with the blue line in Fig.~\ref{fig:drawback}, resembles more the ground truth, and the overall deviation is within $2.3\%$.
Moreover, the de-noised sequence tends to turn back to the ground truth at both ends, which is in contrast to the previous tendency as a result of de-noising with $H$.

\subsection{Low-rank approximation}
For the matrix $X$ defined in Eq.~(\ref{eq:pcm}), the Eckart-Young theorem has asserted that it can be decomposed to (at most) $m$ rank-one matrices~\cite{eckart_approximation_1936}:
\begin{equation}
  X = \sum_{i=0}^{m-1}s_i \mathbf{u}_i \mathbf{v}_i^{\mathsf{T}},
  \label{eq:svd-1}
\end{equation}
where the singular values $s_i$ are non-negative and ordered non-increasingly; the left singular vectors $\mathbf{u}_i\in\mathbb{R}^{m\times1}$ are orthonormal; and similarly for the right singular vectors $\mathbf{v}_i\in\mathbb{R}^{n\times1}$.
When expressed with matrix notation, Eq.~(\ref{eq:svd-1}) leads to the canonical form of SVD:
\begin{equation}
  X = U\Sigma V^{\mathsf{T}},
  \label{eq:svd}
\end{equation}
where $\Sigma$ is an $m\times m$ diagonal matrix composed of $s_i$ as the non-zero elements; $U$ is an $m\times m$ matrix composed of $\mathbf{u}_i$ as the column vectors, while $V$ is an $n\times m$ matrix composed of $\mathbf{v}_i$.

The physical interpretation for $s$ is that each element describes the amount of energy distributed in the corresponding component.
Specifically, the total energy of $x$ is proportional to the squared Frobenius norm of $X$:
\begin{equation}
  \lVert X \rVert^2_F = \sum_{i=0}^{m-1}\sum_{j=0}^{n-1}X_{ij}^2 = m\sum_{i=0}^{n-1}x_i^2.
  \label{eq:f-norm-1}
\end{equation}
Meanwhile, by Eq.~(\ref{eq:svd}), $\lVert X \rVert^2_F$ can be written as
\begin{equation}
  \lVert X \rVert^2_F = \operatorname{trace}\left( X^{\mathsf{T}}X \right) = \operatorname{trace}\left( \Sigma^2 \right) = \sum_{i=0}^{m-1}s_i^2.
  \label{eq:f-norm-2}
\end{equation}

\begin{figure*}
  \includegraphics[width=.9\textwidth]{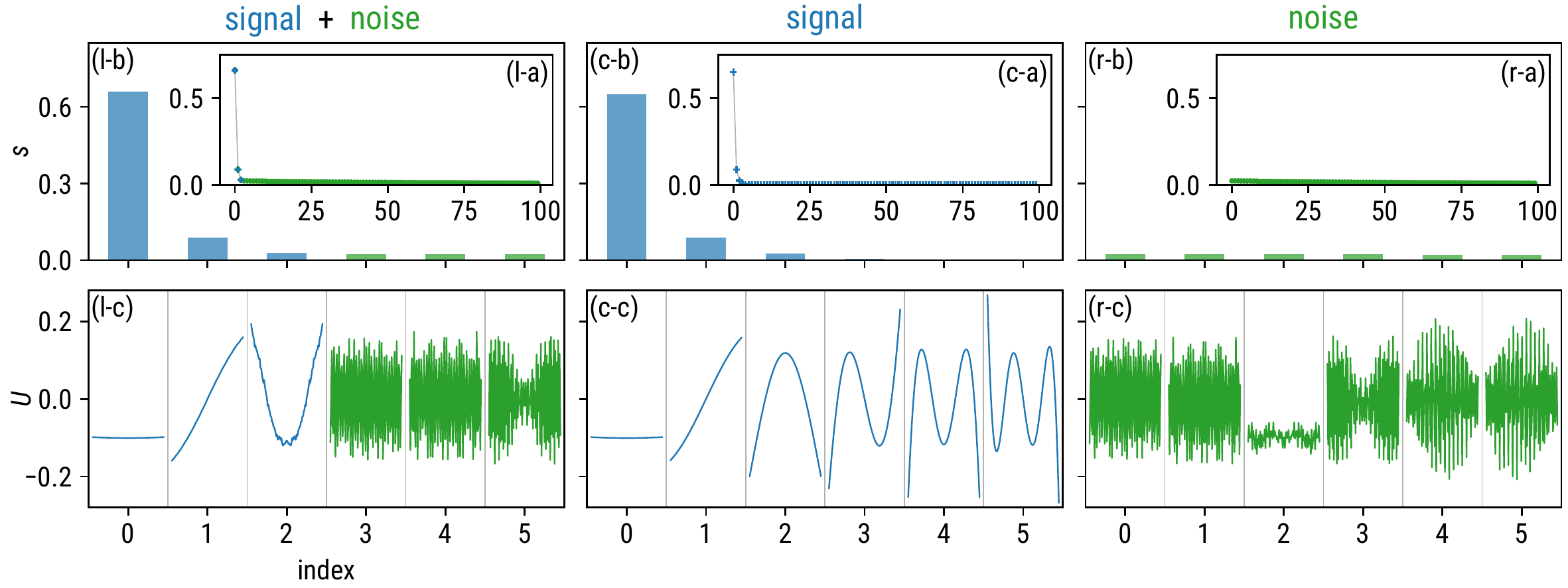}
  \caption{
    Typical examples of SVD components in the case of noise-corrupted signal (left), sole signal (center), and sole noise (right).
    (a) Full spectra of singular values in a non-increasing order, where the blue crosses belong to the signal and the green dots belong to the noise.
    (b) Zoomed spectra focusing on the largest six singular values.
    (c) The corresponding left singular vectors.
    \label{fig:character}
  }
\end{figure*}

Since $s$ is non-negative and non-increasing, by selecting the largest $r$ singular values and setting the remaining $s_i$ to zero, the resultant matrix $A$ is the best rank-$r$ approximation of $X$ in terms of minimizing the approximation error energy~\cite{eckart_approximation_1936}.
This sets the theoretical basis for the data de-noising method proposed here.
In practice, most of the singular values are close to zero [see Fig.~\ref{fig:character}(l-a) for illustration], which can be attributed to the noise.
Therefore, $X$ can be well approximated with a few predominant signal components only.

That said, one question remains to be answered: which index separates the signal components from the noise ones?
Despite different solutions in literature, it is certainly impossible to agree on a universal approach that fits in every realistic situation and fulfills every practical requirement.

Nevertheless, a majority of the methods choose to look into the intrinsic structure of the singular values, or its variants like singular entropy~\cite{yang_development_2003,zhao_selection_2011}, to search for the ``elbow point'' in Fig.~\ref{fig:character}(l-a).
To name a few, one can inspect the first-order difference~\cite{hassanpour_timefrequency_2008,zhao_selection_2011}, the first-order quotient~\cite{cong_short-time_2013,jiang_study_2015}, or simply select the intersection of two extrapolated lines by eye~\cite{schanze_compression_2018}.
All those methods depend on extra deciding parameters, which are usually specified \textit{ad hoc} and vary from case to case.
Consequently, it is rather difficult to implement them in an automated manner to de-noise a large batch of sequences.
What is worse, the smoothed sequence is prone to subjective bias, and thus may vary from person to person.

A more advanced method seeks to approximate $X$ when regularized by the nuclear norm of the low-rank approximator~\cite{furnival_denoising_2017,candes_exact_2009}.
To select an appropriate regularization parameter, this method relies on various unbiased risk estimators according to the prior knowledge on the noise statistics.
Other sophisticated methods based on neural network dictionary learning require considerable training data, and often involve many iterations to solve non-linear minimization problems~\cite{al-zaben_extraction_2006,gong_improved_2017}.
Apparently, those methods are computationally rather intense.

Here, we propose to tackle this challenge by inspecting the left singular vectors.
As can be seen in Fig.~\ref{fig:character}(l-c), the first three belonging to the signal are quite smooth, whereas the noise vectors are very wiggly.
Moreover, there exists a drastic behavioral change at the critical index, which separates the signal from the noise.
This observation is further confirmed by Figs.~\ref{fig:character}(c-c) and \ref{fig:character}(r-c), where the signal and the noise are individually inspected.
To characterize such an impression, a quantity named Normalized Mean Total Variation (NMTV) is defined.
For a given sequence $w$ of length $m$, the NMTV $\xi$ is calculated as
\begin{equation}
  \xi = \frac{\sum_{i=0}^{m-2}\left| w_i-w_{i+1} \right|}{(m-1)(w_{\mathrm{max}}-w_{\mathrm{min}})}.
  \label{eq:nmtv}
\end{equation}

Obviously, $\xi$ is between zero and one, where zero means no variation and one corresponds to a zigzag pattern.
A larger NMTV indicates that the sequence is more oscillatory.
Hence, the NMTV is able to describe the smoothness of the sequence.
Since the noise is usually wiggly, its SVD components can be discriminated by thresholding on the NMTVs of $\mathbf{u}_i$.
Numerical experiments have revealed that a universal threshold of $0.1$ should be suitable in most, if not all, cases.

\subsection{Signal reconstruction}
Having selected the first $r$ SVD components, the low-rank approximation $A$ of $X$ is
\begin{equation}
  A = \sum_{i=0}^{r-1}s_i \mathbf{u}_i \mathbf{v}_i^{\mathsf{T}}.
  \label{eq:low-rank-approx}
\end{equation}
In general, $A$ is no longer partially circulant.
Consequently, cyclic anti-diagonal averaging is applied to reconstruct the smoothed sequence $\hat{a}$ from $A$:
\begin{equation}
  \hat{a}_i = \operatorname*{mean}_{j+k\equiv i\!\pmod{n}}\left( A_{jk} \right),\quad 0\leq i<n.
  \label{eq:anti-diag-mean}
\end{equation}

\subsection{Practical considerations}
So far, the choice of $m$ has not been addressed.
It is nevertheless worth noting that when $m=n$, i.e., $X$ is completely circulant, the method proposed here is equivalent to the Fourier thresholding~\cite{huang_noise_1974}.
Since any circulant matrix is diagonalizable by the unitary discrete Fourier transform matrix, its singular values obtained by SVD are in fact the modulus of the Fourier coefficients of $x$~\cite{karner_spectral_2003}.

However, to select a larger $m$ does not necessarily result in a better smoothing effect, as the signal energy is diluted by excessive singular values.
A relatively small $m$ is not much helpful either due to an insufficient number of elements for the cyclic anti-diagonal averaging.
Although in practice the choice of $m$ may be subject to the specific purpose of the de-noising problem, one can generally decide on an optimal $m$ which minimizes the NMTV of the de-noised sequence to attain the smoothest result.
Numerical experiments have suggested an empirical rule to achieve an optimal de-noising effect: $0.1n<m<0.4n$.

\begin{figure}
  \includegraphics[width=.9\columnwidth]{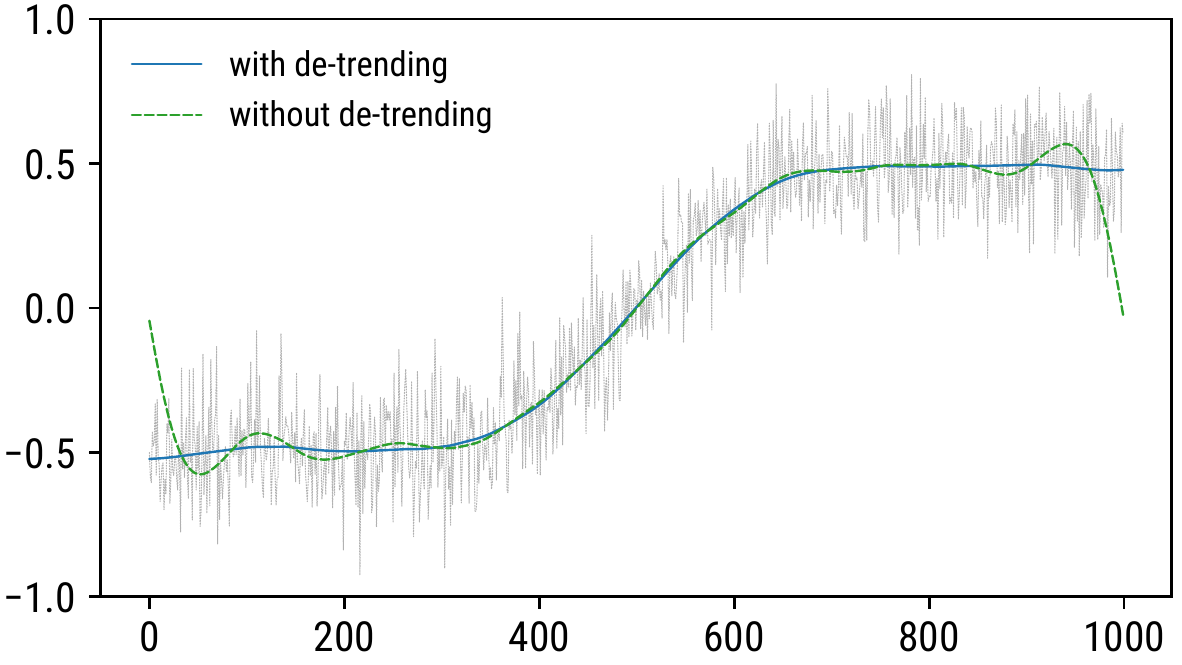}
  \caption{
    Illustration of the ringing artifact in the smoothed data.
    The blue line is obtained with linear de-trending, whereas the green line without.
    \label{fig:detrend}
  }
\end{figure}

Sometimes, it will be unfortunate to notice a ringing artifact in the smoothed data, in particular when the signal has a significant gap between its two ends (Fig.~\ref{fig:detrend}).
This should be owing to the same reason as for the Gibbs phenomenon in Fourier transform~\cite{gibbs_fouriers_1898}.
As a cure for the artifact, a linear trend, if at all, in the signal should be removed to close the gap before de-noising.
Whether the de-trending is practically necessary is judged by a comparison between the signal gap $\Delta$ and the noise standard deviation $\sigma$.

Unfortunately, neither of them are known, and hence they must be estimated from $x$.
Despite the wiggles at both ends of $x$, the respective signal end-values can be recovered somehow by averaging a few neighboring data, based on which $\Delta$ is estimated.
Moreover, $\sigma$ can be estimated from the noise singular values.
By recalling Eqs.~(\ref{eq:add}), (\ref{eq:f-norm-1}), and (\ref{eq:f-norm-2}), it can be written that
\begin{equation}
  \sum_{i=0}^{n-1}a_i^2 + \sum_{i=0}^{n-1}d_i^2 + 2\sum_{i=0}^{n-1}a_id_i = \frac{1}{m}\sum_{i=0}^{r-1}s_i^2 + \frac{1}{m}\sum_{i=r}^{m-1}s_i^2,
\end{equation}
where, due to the cancellation among $d_i$, the summation of the cross terms on the left side contributes rather little.
Therefore, $\sigma$ can be approximated as
\begin{equation}
  \sigma = \sqrt{\frac{\sum_{i=0}^{n-1}d_i^2}{n}} \approx \sqrt{\frac{\sum_{i=r}^{m-1}s_i^2}{mn}}.
  \label{eq:sigma}
\end{equation}

Should $\sigma$ turn out to be less than $\Delta$ after de-noising, a linear trend is significantly present and must be removed.
However, it may happen sometimes that for the de-trended yet noisy sequence, $\sigma$ is still less than $\Delta$, which suggests that the estimate of the latter by averaging is strongly biased by the signal itself.
Under such circumstances, it is advisable to reduce the number of elements for averaging and redo the de-trending.
This adaptive de-trending process will surely terminate, in the worst case, when the averaging number becomes one.

In summary, the algorithm of the proposed data de-noising method is organized as follows.

\begin{enumerate}
  \item Begin with a noisy sequence $x$ and a row number $m$.
  \item Construct a partial circulant matrix $X$ by Eq.~(\ref{eq:pcm}), then apply SVD to it. \label{loop:start}
  \item Calculate the NMTVs of the left singular vectors $\mathbf{u}_i$ by Eq.~(\ref{eq:nmtv}), then select the smallest $r$ that satisfies $\operatorname{NMTV}\left( \mathbf{u}_{r-1} \right)<0.1\leq \operatorname{NMTV}\left( \mathbf{u}_r \right)$ as the approximation rank.
  \item Estimate the signal gap $\Delta$ by averaging a few end-data in $x$, as well as the noise standard deviation $\sigma$ by Eq.~(\ref{eq:sigma}). \label{loop:stop}
  \item If $\sigma<\Delta$, linearly de-trend $x$, then repeat steps~\ref{loop:start}--\ref{loop:stop}; otherwise proceed.
  \item Build a low-rank approximation $A$ by Eq.~(\ref{eq:low-rank-approx}).
  \item End with a smoothed sequence $\hat{a}$ by Eq.~(\ref{eq:anti-diag-mean}).
\end{enumerate}

\section{Examples}
To illustrate the smoothing effect of the proposed method, four synthetic signals of the same length $n=1000$ are adopted for test.
They are intentionally selected to embody different data traits that are frequently seen in practice.
Specifically, their synthetic formulas are listed below, with $0\leq i <1000$.

\begin{description}
  \item[periodic signal]
    \begin{equation}
      a_i = \sin(0.004\pi i) - \sin^3(0.008\pi i) + \cos\left[\sin(0.008\pi i)\right],
    \end{equation}
    which contains exactly $2$ periods.
  \item[asymmetric peak]
    \begin{equation}
      a_i = \operatorname{erfc}(3-0.01i)\cdot e^{-0.01i},
    \end{equation}
    where $\operatorname{erfc}(z)=1 - \pi^{-1/2}\left(\int_{-z}^{z}e^{-t^2}dt\right)$ is the complementary error function.
  \item[multiple peaks on a ramp]
    \begin{equation}
      a_i = e^{-(0.02i-6)^2} + \left[ 2 + (0.04i-28)^2 \right]^{-1} + e^{-0.001i},
    \end{equation}
    which contains a Gaussian peak and a Lorentzian peak.
  \item[chaotic dynamics]
    It is well known that a Duffing oscillator will undergo chaos under certain conditions~\cite{duffing_erzwungene_1918}.
    Here, we adopt the following parametrization of the Duffing equation:
    \begin{equation}
      \ddot{a} + 0.3\dot{a} - a(1-a^2) = 0.5\cos(0.4\pi t),
    \end{equation}
    where $a$ denotes the oscillation amplitude, and the dotted versions denote its derivatives with respect to time $t$.
    With the initial value $(a, \dot{a})_0 = (1, 0)$, the differential equation is numerically solved by the Runge-Kutta fourth-order method in an interval of $0\leq t<50$ with a resolution of $0.05$.
\end{description}

The synthetic signals are then contaminated by adding respectively two types of noise, namely uniform and normal, to produce noisy sequences.
The amount of the contamination is controlled by the Signal-to-Noise Ratio (SNR) ranging from $0$ to $20$~dB, which is defined as:
\begin{equation}
  \rho_{(\mathrm{dB})} = 10\log_{10}\left( \frac{\langle a^2 \rangle - \langle a \rangle^2}{\langle d^2 \rangle - \langle d \rangle^2} \right),
\end{equation}
where $\langle \cdot \rangle$ denotes arithmetic mean.
It is clear that a smaller $\rho$ will lead to a more wiggly sequence.

A fixed row number $m=250$ is selected for all the smoothing tests.
The de-noising goodness is assessed by the Normalized Root-Mean-Square Deviation (NRMSD) of the smoothed sequence $\hat{a}$ from the signal $a$, which is defined as
\begin{equation}
  \delta_{(\%)} = 100\sqrt{\frac{\langle\left( \hat{a}-a \right)^2\rangle}{\langle a^2 \rangle - \langle a \rangle^2}}.
\end{equation}
A smaller $\delta$ indicates a greater power to reveal the ground truth.

\begin{figure*}
  \includegraphics[width=.9\textwidth]{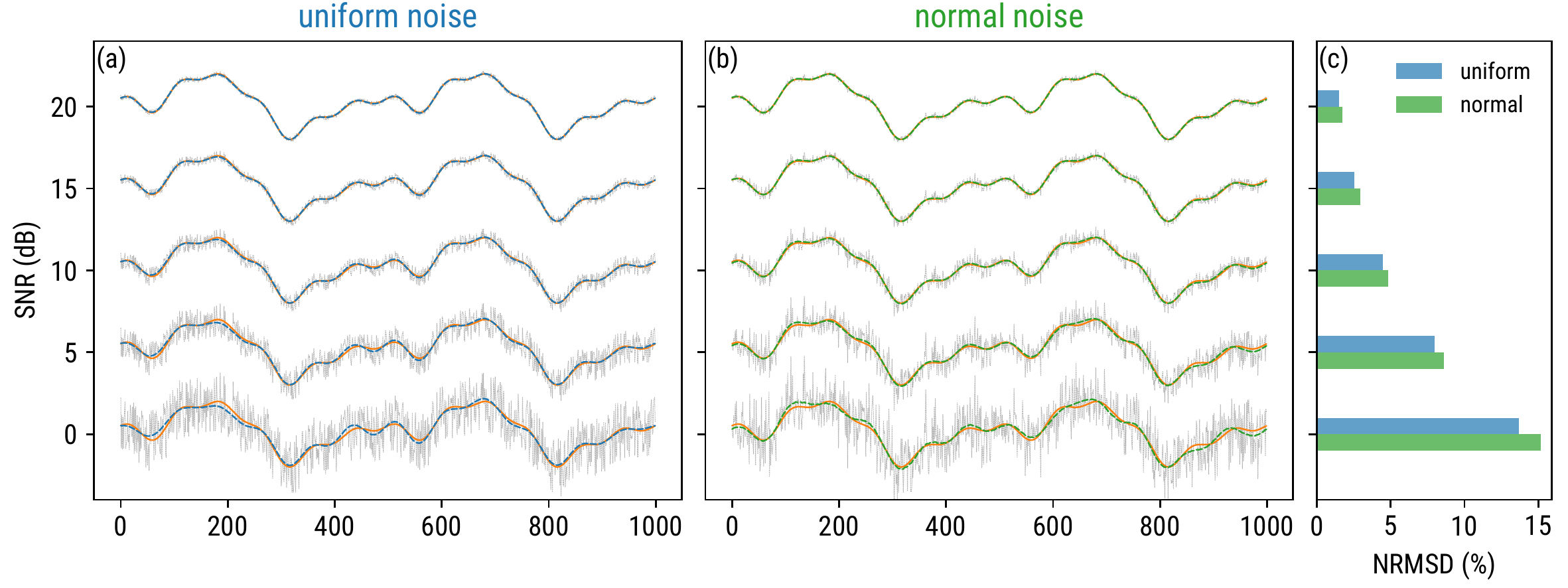}
  \caption{
    Illustration of the smoothing effect on the noise-corrupted periodic signals with various SNRs.
    (a) The noise obeys a uniform distribution.
    The orange line is the ground truth, while the blue line is the de-noised approximation.
    (b) Same as (a), except the noise is normally distributed.
    (c) The resultant NRMSDs obtained in different cases.
    \label{fig:periodic}
  }
\end{figure*}

\begin{figure*}
  \includegraphics[width=.9\textwidth]{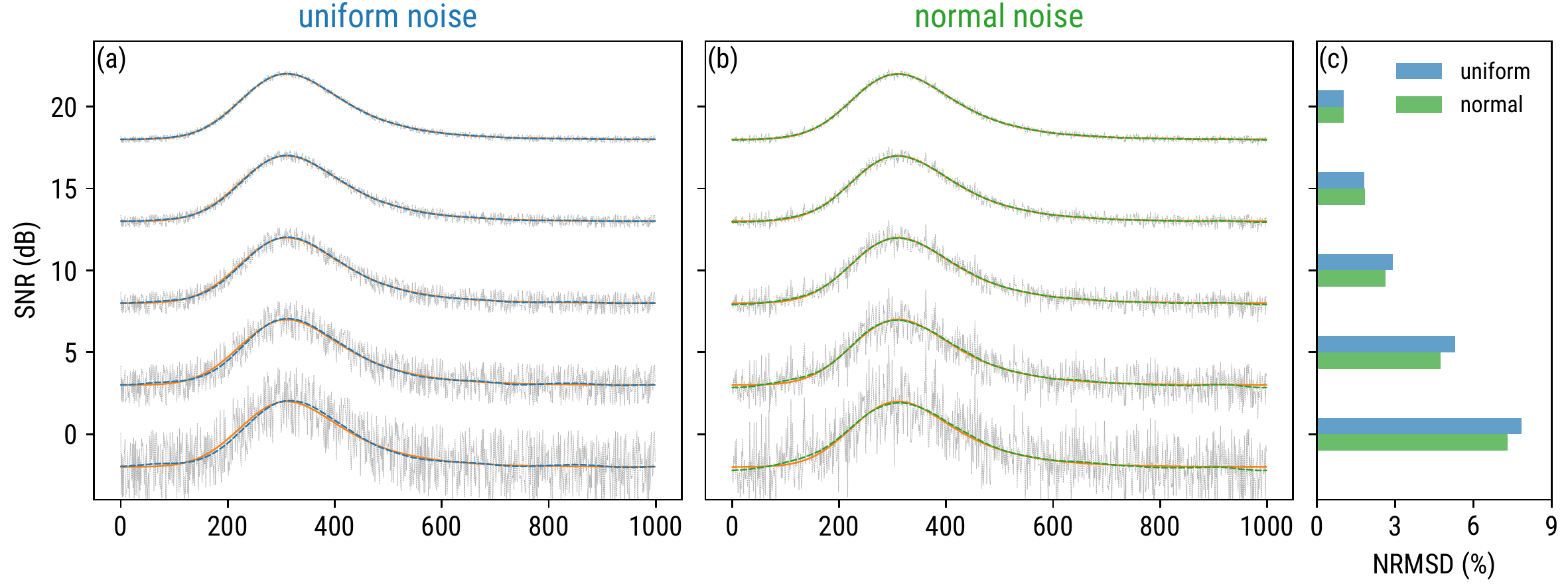}
  \caption{
    Same as Fig.~\ref{fig:periodic}, but for an asymmetric peak.
    \label{fig:asymmetric}
  }
\end{figure*}

\begin{figure*}
  \includegraphics[width=.9\textwidth]{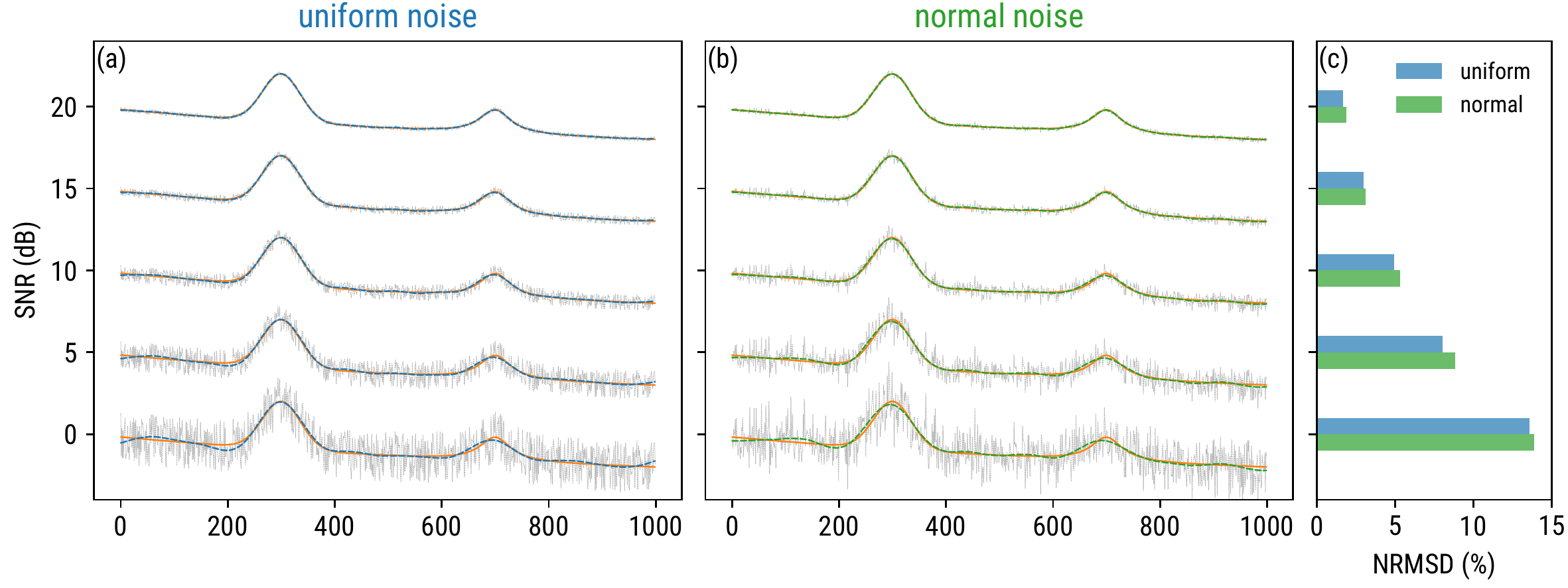}
  \caption{
    Same as Fig.~\ref{fig:periodic}, but for a Gaussian peak (left) and a Lorentzian peak (right) on a ramp.
    \label{fig:multipeak}
  }
\end{figure*}

\begin{figure*}
  \includegraphics[width=.9\textwidth]{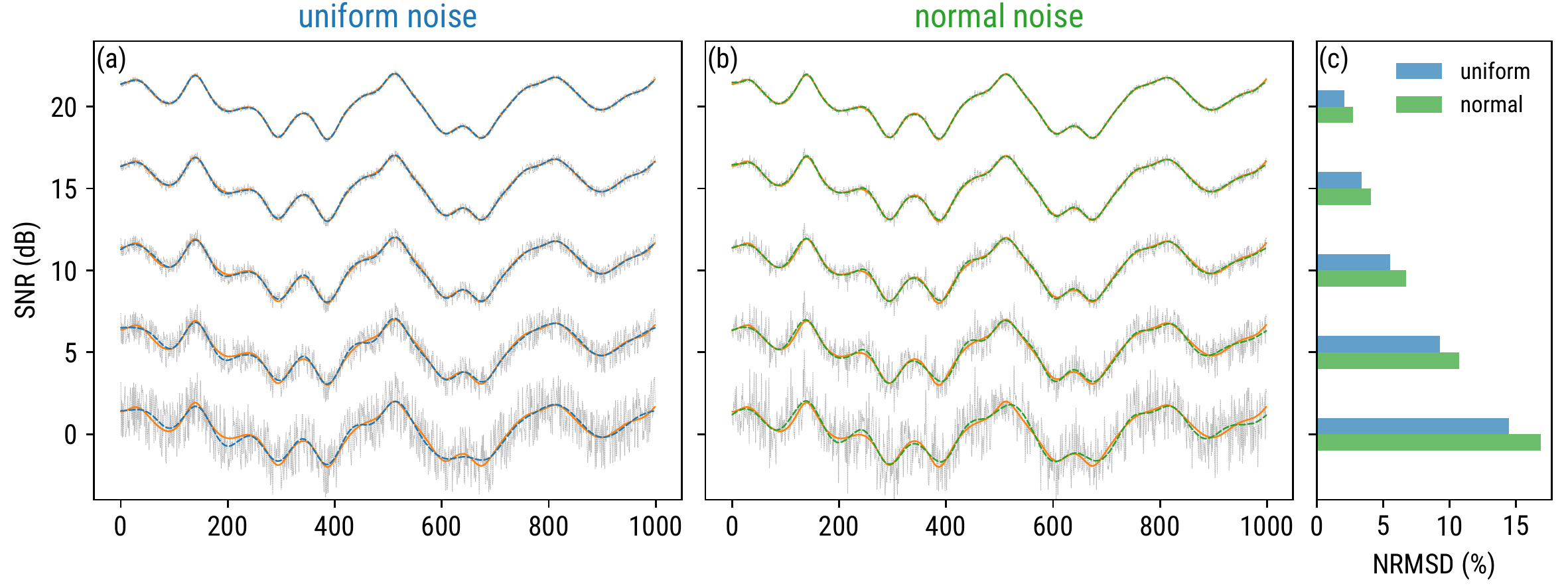}
  \caption{
    Same as Fig.~\ref{fig:periodic}, but for the amplitude of a Duffing oscillator.
    \label{fig:chaotic}
  }
\end{figure*}

The test results are shown in Figs.~\ref{fig:periodic}--\ref{fig:chaotic}.
It can be seen that the proposed method is well able to smooth the noisy sequences, regardless of the underlying signals, even when the SNR is down to $0$~dB.
The achieved NRMSDs are quite similar for both types of noise, although the normal noise usually results in a slightly larger NRMSD, which may be owing to its larger kurtosis.

\section{Applications}
Owing to its non-parametric nature, the method presented here is promising to find its applications in vast data de-noising scenarios.
Furthermore, by allowing for a manual selection of the approximation rank, the proposed method will be a powerful data analysis technique not just for de-noising.
In what follows, three data analysis challenges in precision storage-ring mass spectrometry are presented to showcase its diverse applications in reality~\cite{franzke_mass_2008}.

\subsection{Pulse timing} \label{sec:timing}
\begin{figure}
  \includegraphics[width=.9\columnwidth]{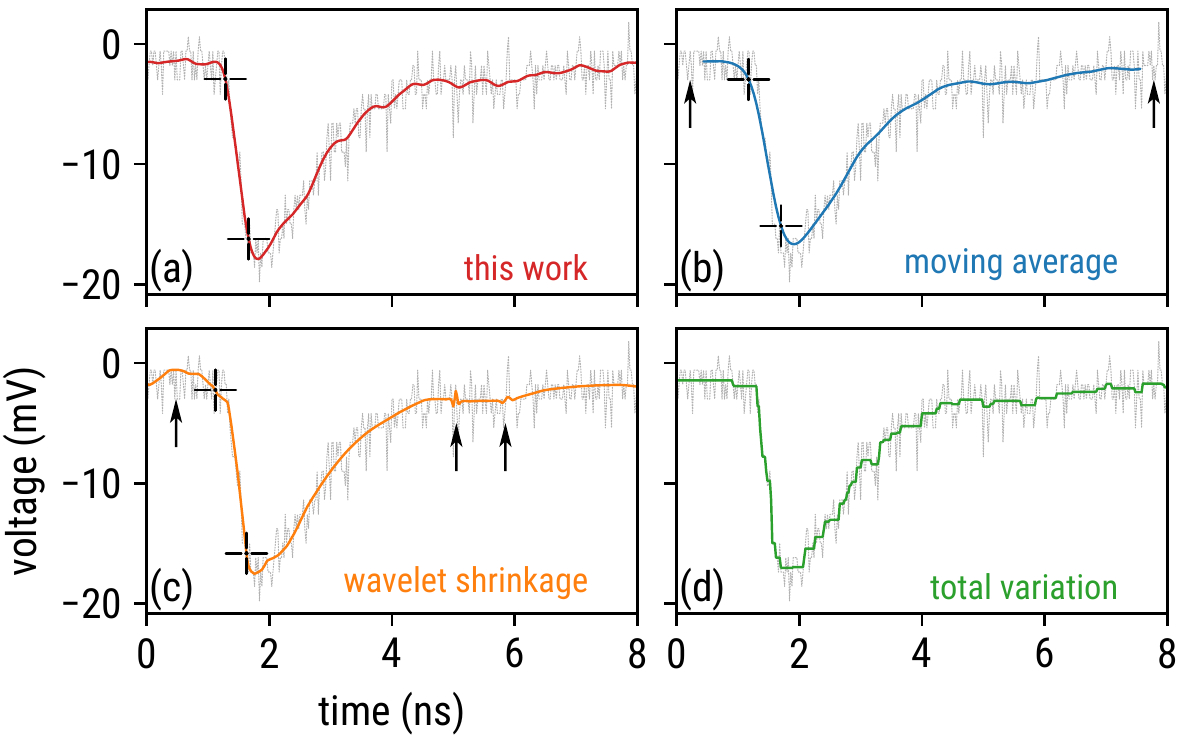}
  \caption{
    A zoomed pulse recorded by a time-of-flight detector, which is de-noised by (a) the proposed method in this work, (b) moving average, (c) wavelet shrinkage, and (d) total variation.
    The crosses on the leading edge mark the $90\%$ to $10\%$ interval used to calculate the fall time.
    See Sec.~\ref{sec:timing} for the detail.
    \label{fig:timing}
  }
\end{figure}

In the isochronous mass spectrometry~\cite{hausmann_first_2000}, a relativistic charged particle periodically circulates in a vacuum environment due to the confinement by magnetic field~\cite{xu_accurate_2013}.
A time-of-flight detector consisting of an ultra thin carbon foil is placed in the vacuum chamber to register every passage of the particle, which is signaled by a sharp voltage pulse~\cite{mei_high_2010}.
Fig.~\ref{fig:timing} shows a zoomed example of such a pulse.
A precise determination of the pulse onset is critically important for a high-precision ($<1$ ppm) mass measurement~\cite{tu_direct_2011,zhang_mass_2012,xu_identification_2016}.
Unfortunately, due to thermal noise and quantization error, in particular for weak pulses, the pulse leading edge is distorted, which limits the timing accuracy.
Furthermore, the voltage uncertainty in pulse samples also degrades the timing precision, since the leading edge can only have a non-zero fall time, which is the duration for the pulse to dive from $90\%$ to $10\%$ of its full depth.
In general, a smaller fall time will result in a better timing precision.
As for the timing accuracy, it can be improved by employing a competent data de-noising technique.

With the same pulse of length $n=401$ in Fig.~\ref{fig:timing}, we have compared four de-noising methods of non-parametric type, namely the one proposed here, moving average, wavelet shrinkage, and total variation.
The NMTVs of their de-noised data, which are shown in Table~\ref{tab:compare}, are controlled to be almost the same to ensure comparable smoothness.
Afterwards, the fall time is measured for each smoothed sequence, except for the total variation since the staircase effect unfeasibly allows for a clear definition of falling interval.
In what follows, the technical implementing details are presented.

\begin{table}[htbp]
  \caption{\label{tab:compare}Characteristics of de-noised data by various methods.}
  \begin{ruledtabular}
    \begin{tabular}{lcc}
      Method & NMTV (\textperthousand) & Fall time (ns)\\
      \hline
      this work & 5.70 & \textbf{0.38} \\
      moving average & 5.69 & 0.53 \\
      wavelet shrinkage & 5.71 & 0.51 \\
      total variation & 5.70 & --- \\
    \end{tabular}
  \end{ruledtabular}
\end{table}

For the proposed method, a row number $m=105$ is determined.
As for the moving average, a Gaussian kernel $g$ of length $45$ and standard deviation $9$ is employed:
\begin{equation}
  g_i \propto e^{-i^2/162}, \quad -22\leq i \leq22.
\end{equation}
When applying the wavelet shrinkage, the \texttt{Symlets-4} is adopted to perform the wavelet decomposition.
Afterwards, the wavelet coefficients are thresholded on $3.406$ using the non-negative garrote scheme~\cite{breiman_better_1995,gao_wavelet_1998}.
The wavelet transform is carried out in \texttt{Python} with the \texttt{PyWavelets} package~\cite{lee_pywavelets_2006}.
Lastly, the total variation is based on the algorithm for one-dimensional data in Ref.~\cite{condat_direct_2013}, and computed with the \texttt{C} code available in Ref.~\cite{condat_total_2017}.
The regularization parameter for total variation is selected to be $2.89$.

The de-noised results by these four methods are shown in Fig.~\ref{fig:timing}, and the obtained fall times are listed in Table~\ref{tab:compare}.
It is found that the proposed method produces the smallest fall time, which can in principle lead to the best timing precision.
In contrast, the wide kernel of moving average not only leaves apparent vacancy in the de-noised data at both ends, as pointed by arrows in Fig.~\ref{fig:timing}(b), but also smears out the leading edge.
The wavelet shrinkage superfluously causes a leading bump and trailing spikes, as pointed by arrows in Fig.~\ref{fig:timing}(c), where the former also degrades the fall time.
As for the total variation, the disturbing staircase effect, as shown in Fig.~\ref{fig:timing}(d), makes it improper for accurate and precise pulse timing.
It can therefore be concluded that by means of the proposed method, timing of weak pulses has become feasible with satisfactory accuracy and precision, thus relaxing the saturation constraint of the detector~\cite{zhang_timing_2014}.

\subsection{Pre-whitening}
In the Schottky mass spectrometry~\cite{litvinov_mass_2005}, a dedicated resonator is employed to detect a revolving particle in the storage ring by sensing its electromagnetic radiation~\cite{nolden_fast_2011}.
Owing to the periodic motion of the particle, the power spectral density of its Schottky signal detected by the resonator peaks at each harmonic of the particle's revolution frequency~\cite{chattopadhyay_fundamental_1985}.
Among those harmonics, the one in the resonant range of the resonator is band-pass filtered and mixed down to the base band for data acquisition~\cite{trageser_new_2015}.

\begin{figure}
  \includegraphics[width=.9\columnwidth]{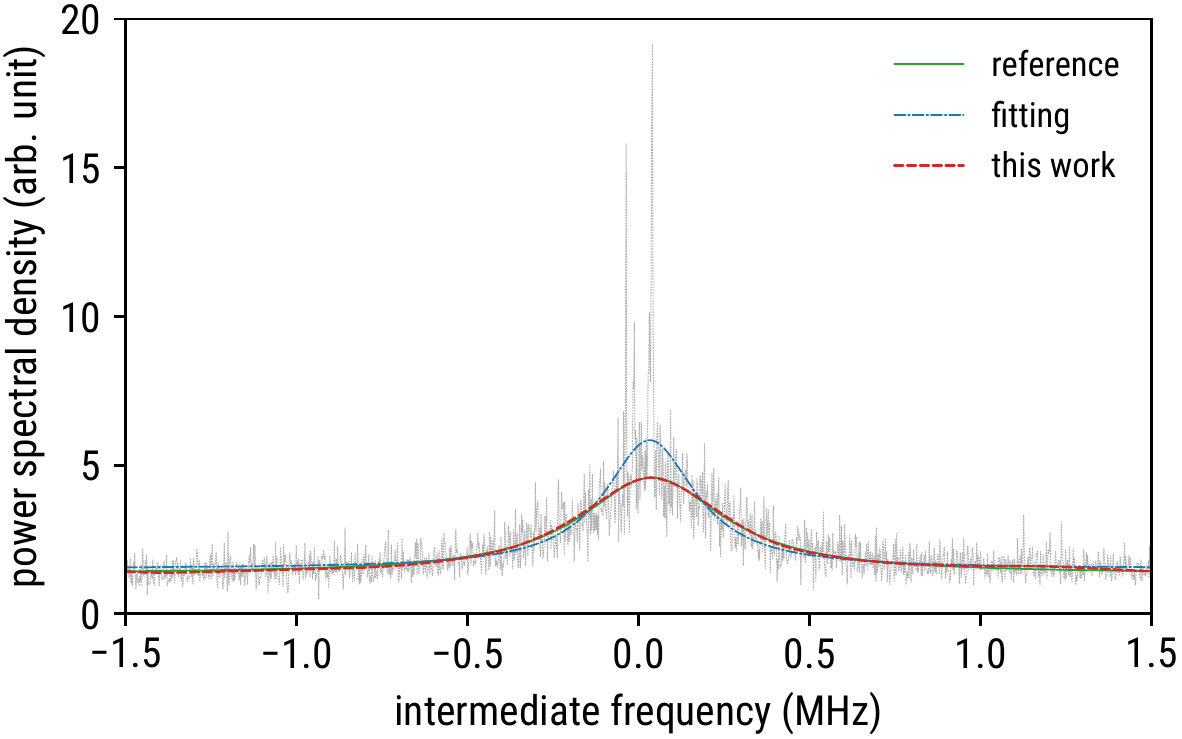}
  \caption{
    Background estimation of a Schottky spectrum with a sharp peak on top of it.
    The blue line is a direct fitting of a Lorentzian function, whereas the red line is obtained by the proposed method in this work.
    The green line acts like a reference.
    \label{fig:prewhitening}
  }
\end{figure}

Fig.~\ref{fig:prewhitening} shows an example of a peak harmonic on top of a resonant background.
In certain cases, the integral peak area is of the experimental interest, from which the particle number can be inferred~\cite{litvinov_beta_2011,chattopadhyay_fundamental_1985}.
This is a prior step for, e.g., the measurement of the particle's lifetime~\cite{tu_first_2018}, or the calibration of the particle's position~\cite{chen_accuracy_2015}.
Therefore, a proper deduction of the background is of the experimental importance.

Although the Lorentzian shape of the background is well understood~\cite{chen_non-interceptive_2015}, it is improper to directly apply fitting to the spectrum as the peak will strongly deviate the result (see blue line in Fig.~\ref{fig:prewhitening}).
In contrast, the method proposed here with a bit of adaptations can elegantly eliminate the peak interference.
First recall that the Fourier coefficients of any noise are zero-mean circular complex Gaussian~\cite{schoukens_modeling_1986}.
Consequently, the power spectral density of the noise, which is the squared modulus of the coefficients, obeys a chi-squared distribution with $2$ degrees of freedom.
Its probability density function is in fact a negative exponent, which happens to be the same distribution as a speckle noise obeys~\cite{goodman_statistical_1975}.
Since the speckle noise is usually modeled as a multiplicative noise~\cite{arsenault_image_1983}, it is analogous to conclude that the noise in the Schottky spectrum is also multiplicative.

Therefore, a logarithmic transform should first be applied to the spectrum so as to use the additive model in Eq.~(\ref{eq:add}).
The transformed spectrum of length $n=1639$ is then embedded into a partial circulant matrix of $m=290$ rows.
This time only the first $2$ components are taken to reconstruct the background.
As can be seen in Fig.~\ref{fig:prewhitening}, the resultant approximation of red line is in excellent agreement with the reference of green line, which is obtained by fitting a ``bald'' spectrum when particles are purged from the storage ring.

The proposed method hence opens up a possibility to deduct the resonant background \textit{in situ}.
It will be no longer necessary to interrupt an ongoing experiment just for the reference measurements, which increases effective on-target beam time.
This is in particular useful in case the yields of aimed particles are extremely low.

\subsection{Peak detection}
\begin{figure}
  \includegraphics[width=.9\columnwidth]{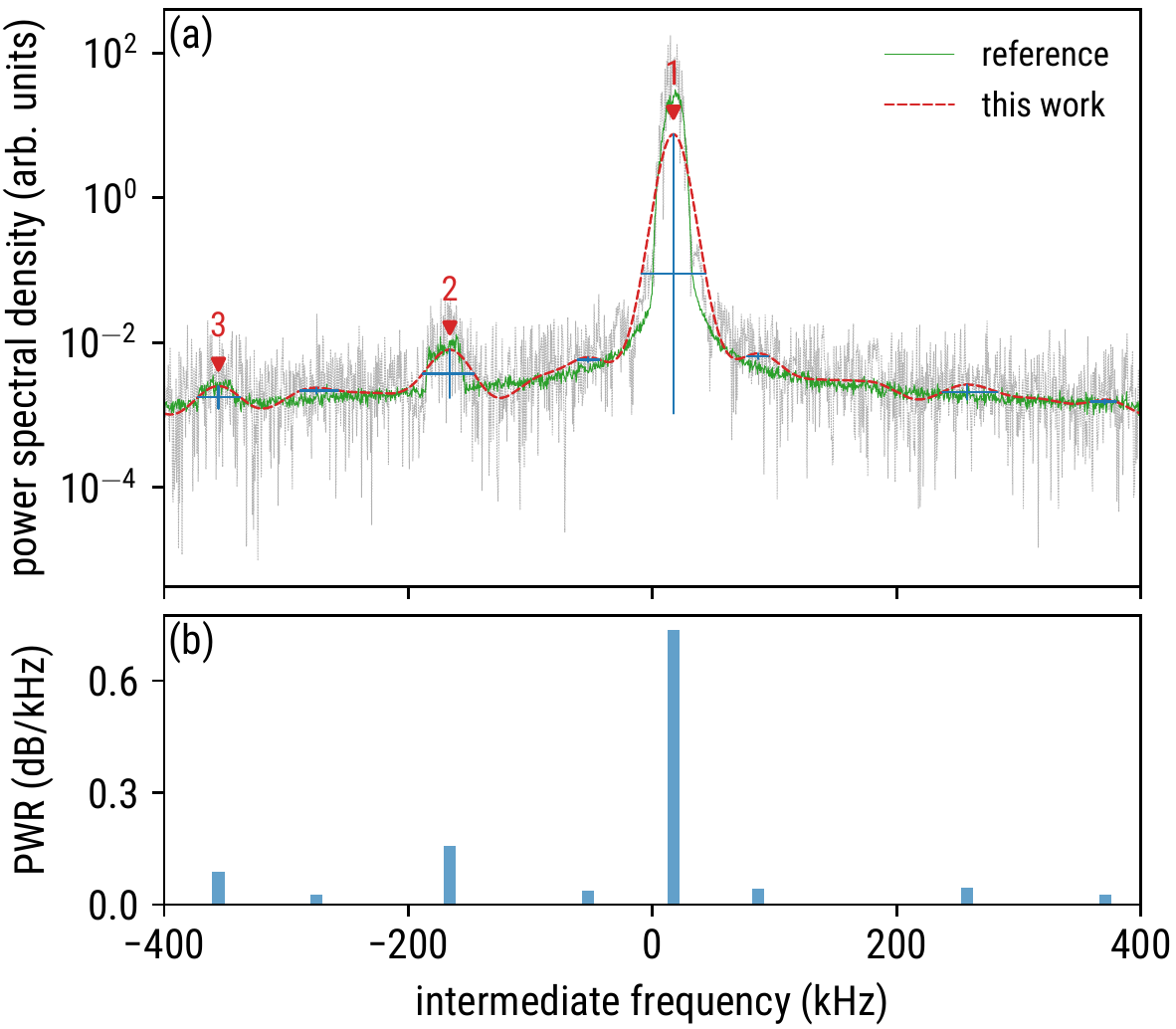}
  \caption{
    Weak peak detection in a Schottky spectrum by the proposed method in this work.
    (a) The red line is the de-noised spectrum, and the green line, which is obtained by averaging $100$ similar spectra, provides a reference.
    All local maxima in the red spectrum are marked with blue crosses, which also respectively indicate their prominences (vertical extent) and widths (horizontal extent).
    The detected peaks are labelled, in a descending order of PWR, as 1, 2, and 3.
    (b) The PWR of all eight peak candidates, among which the largest three are considered as real peaks associated with particles.
    \label{fig:peakdec}
  }
\end{figure}

Another common challenge in Schottky mass spectrometry is the detection of weak peaks.
Fig.~\ref{fig:peakdec}(a) shows an example of such a case.
Apart from the predominant peak in the center, it is impossible to unambiguously detect other peaks in the noisy spectrum.
A normal solution is to continuously collect a number of spectra and average them to improve the SNR.
For example, the green line in Fig.~\ref{fig:peakdec}(a) is obtained from $100$-fold averaging, which can barely reveal two weaker peaks on the left side.
An obvious drawback of this approach is the degradation of time resolution, which, in this case, is $100$ times worse.
It is thus unfavorable for the time-resolved Schottky mass spectrometry~\cite{litvinov_precision_2004}, and in particular for the single-particle decay spectroscopy~\cite{kienle_high-resolution_2013}.
Even worse is that sometimes it is unfeasible to collect sufficient spectra, because, e.g., the aimed particle may have already decayed~\cite{walker_ilima_2013}.

By using the proposed method, the noisy spectrum of length $n=1639$ is first log-transformed and then embedded into a partial circulant matrix of $m=430$ rows.
A smoothed spectrum is reconstructed from the first $7$ SVD components, which is shown as the red line in Fig.~\ref{fig:peakdec}(a).
Peaks are searched among local maxima in the smoothed spectrum as the candidates.
For each candidate, its prominence and width are subsequently measured, where the prominence is the relative height from the summit to the highest adjacent valley, and the width is the horizontal span at half prominence [see blue crosses in Fig.~\ref{fig:peakdec}(a) for illustration].
Afterwards, the Prominence-to-Width Ratio (PWR), shown in Fig.~\ref{fig:peakdec}(b), is calculated for each candidate to characterize its significance.
Based on this parameter, in total three peaks are decisively identified with the largest PWRs and labelled with 1, 2, and 3 in Fig.~\ref{fig:peakdec}(a).
The remaining candidates are only considered as bumps due to their insignificant PWRs.
It is clear in Fig.~\ref{fig:peakdec}(a) that the detected peaks perfectly align with those in the averaged spectrum, and, moreover, with a $100$-fold improvement of time resolving power.

Once the locations of weak peaks are coarsely determined from the smoothed spectrum, they can subsequently be transferred to a dedicated peak-finding algorithm as the required prior knowledge to refine the result~\cite{chen_band-limited_2017}.
What is more, the proposed method can be deployed to a real-time task to fast track particles in a dynamic process such as orbital-electron capture~\cite{litvinov_measurement_2007} or bound-state $\beta^-$ decay~\cite{bosch_observation_1996} on an event-by-event basis.

\section{Conclusion}
An SVD-based one-dimensional data de-noising scheme has been proposed.
Owing to its non-parametric nature, it works for any additive noise, as well as multiplicative noise after logarithmic transform.
By construction of a partial circulant matrix, the proposed method well balances computational simplicity and smoothing efficiency.
The proposed NMTV criterion to select the optimal approximation rank allows for an integration into automated processes, which is a compelling feature for certain applications in reality.
With the real-world data, the proposed method has proved its strong competitiveness among other non-parametric de-noising methods such as moving average, wavelet shrinkage, and total variation.
Furthermore, a few interesting applications of the method other than de-noising were also addressed, and its competence has been demonstrated with cases in precision storage-ring mass spectrometry.
Lastly, a \texttt{Python} implementation of the proposed method is provided and placed in the public domain~\cite{chen_generic_2019}.

\begin{acknowledgments}
  This work was partially supported by the CSC and DAAD under the Project Based Personnel Exchange Program (No.\ 57389367), by the Key Research Program of Frontier Sciences of CAS (No.\ QYZDJ-SSW-S, No.\ XDPB09), by the Chinese National Key Program for Science and Technology R\&D (No.\ 2016YFA0400504, No.\ 2018YFA0404400), and by the Major State Basic Research Development Program of China (No.\ 2013CB834401).
  Y.A.L.\ acknowledges the support by the CAS President's International Fellowship Initiative (No.\ 2016VMA043), by the European Research Council (ERC) under the Horizon 2020 Research and Innovation Program (No.\ 682841 ``ASTRUm'').
  Y.H.Z.\ acknowledges the support by the ExtreMe Matter Institute (EMMI) at GSI.
\end{acknowledgments}


\begin{thebibliography}{71}%
\makeatletter
\providecommand \@ifxundefined [1]{%
 \@ifx{#1\undefined}
}%
\providecommand \@ifnum [1]{%
 \ifnum #1\expandafter \@firstoftwo
 \else \expandafter \@secondoftwo
 \fi
}%
\providecommand \@ifx [1]{%
 \ifx #1\expandafter \@firstoftwo
 \else \expandafter \@secondoftwo
 \fi
}%
\providecommand \natexlab [1]{#1}%
\providecommand \enquote  [1]{``#1''}%
\providecommand \bibnamefont  [1]{#1}%
\providecommand \bibfnamefont [1]{#1}%
\providecommand \citenamefont [1]{#1}%
\providecommand \href@noop [0]{\@secondoftwo}%
\providecommand \href [0]{\begingroup \@sanitize@url \@href}%
\providecommand \@href[1]{\@@startlink{#1}\@@href}%
\providecommand \@@href[1]{\endgroup#1\@@endlink}%
\providecommand \@sanitize@url [0]{\catcode `\\12\catcode `\$12\catcode
  `\&12\catcode `\#12\catcode `\^12\catcode `\_12\catcode `\%12\relax}%
\providecommand \@@startlink[1]{}%
\providecommand \@@endlink[0]{}%
\providecommand \url  [0]{\begingroup\@sanitize@url \@url }%
\providecommand \@url [1]{\endgroup\@href {#1}{\urlprefix }}%
\providecommand \urlprefix  [0]{URL }%
\providecommand \Eprint [0]{\href }%
\providecommand \doibase [0]{https://doi.org/}%
\providecommand \selectlanguage [0]{\@gobble}%
\providecommand \bibinfo  [0]{\@secondoftwo}%
\providecommand \bibfield  [0]{\@secondoftwo}%
\providecommand \translation [1]{[#1]}%
\providecommand \BibitemOpen [0]{}%
\providecommand \bibitemStop [0]{}%
\providecommand \bibitemNoStop [0]{.\EOS\space}%
\providecommand \EOS [0]{\spacefactor3000\relax}%
\providecommand \BibitemShut  [1]{\csname bibitem#1\endcsname}%
\let\auto@bib@innerbib\@empty
\bibitem [{\citenamefont {Hyndman}(2011)}]{hyndman_moving_2011}%
  \BibitemOpen
  \bibfield  {author} {\bibinfo {author} {\bibfnamefont {R.~J.}\ \bibnamefont
  {Hyndman}},\ }\bibfield  {title} {\bibinfo {title} {Moving {Averages}},\ }in\
  \href {https://doi.org/10.1007/978-3-642-04898-2_380} {\emph {\bibinfo
  {booktitle} {International {Encyclopedia} of {Statistical} {Science}}}},\
  \bibinfo {editor} {edited by\ \bibinfo {editor} {\bibfnamefont
  {M.}~\bibnamefont {Lovric}}}\ (\bibinfo  {publisher} {Springer},\ \bibinfo
  {address} {Berlin, Heidelberg},\ \bibinfo {year} {2011})\ pp.\ \bibinfo
  {pages} {866--869}\BibitemShut {NoStop}%
\bibitem [{\citenamefont {Clark}(1977)}]{clark_non-parametric_1977}%
  \BibitemOpen
  \bibfield  {author} {\bibinfo {author} {\bibfnamefont {R.~M.}\ \bibnamefont
  {Clark}},\ }\bibfield  {title} {\bibinfo {title} {Non-{Parametric}
  {Estimation} of a {Smooth} {Regression} {Function}},\ }\href
  {https://doi.org/10.1111/j.2517-6161.1977.tb01611.x} {\bibfield  {journal}
  {\bibinfo  {journal} {J. Royal Stat. Soc. B}\ }\textbf {\bibinfo {volume}
  {39}},\ \bibinfo {pages} {107} (\bibinfo {year} {1977})}\BibitemShut
  {NoStop}%
\bibitem [{\citenamefont {Roberts}(1959)}]{roberts_control_1959}%
  \BibitemOpen
  \bibfield  {author} {\bibinfo {author} {\bibfnamefont {S.~W.}\ \bibnamefont
  {Roberts}},\ }\bibfield  {title} {\bibinfo {title} {Control {Chart} {Tests}
  {Based} on {Geometric} {Moving} {Averages}},\ }\href
  {https://doi.org/10.1080/00401706.1959.10489860} {\bibfield  {journal}
  {\bibinfo  {journal} {Technometrics}\ }\textbf {\bibinfo {volume} {1}},\
  \bibinfo {pages} {239} (\bibinfo {year} {1959})}\BibitemShut {NoStop}%
\bibitem [{\citenamefont {Tomasi}\ and\ \citenamefont
  {Manduchi}(1998)}]{tomasi_bilateral_1998}%
  \BibitemOpen
  \bibfield  {author} {\bibinfo {author} {\bibfnamefont {C.}~\bibnamefont
  {Tomasi}}\ and\ \bibinfo {author} {\bibfnamefont {R.}~\bibnamefont
  {Manduchi}},\ }\bibfield  {title} {\bibinfo {title} {Bilateral filtering for
  gray and color images},\ }in\ \href
  {https://doi.org/10.1109/ICCV.1998.710815} {\emph {\bibinfo {booktitle}
  {Proc. {ICCV} '98}}}\ (\bibinfo  {publisher} {Narosa Publishing House},\
  \bibinfo {address} {Bombay, India},\ \bibinfo {year} {1998})\ pp.\ \bibinfo
  {pages} {839--846}\BibitemShut {NoStop}%
\bibitem [{\citenamefont {Buades}\ \emph {et~al.}(2005)\citenamefont {Buades},
  \citenamefont {Coll},\ and\ \citenamefont {Morel}}]{buades_non-local_2005}%
  \BibitemOpen
  \bibfield  {author} {\bibinfo {author} {\bibfnamefont {A.}~\bibnamefont
  {Buades}}, \bibinfo {author} {\bibfnamefont {B.}~\bibnamefont {Coll}},\ and\
  \bibinfo {author} {\bibfnamefont {J.}~\bibnamefont {Morel}},\ }\bibfield
  {title} {\bibinfo {title} {A non-local algorithm for image denoising},\ }in\
  \href {https://doi.org/10.1109/CVPR.2005.38} {\emph {\bibinfo {booktitle}
  {Proc. {CVPR} '05}}},\ Vol.~\bibinfo {volume} {2}\ (\bibinfo {address} {San
  Diego, CA, USA},\ \bibinfo {year} {2005})\ pp.\ \bibinfo {pages}
  {60--65}\BibitemShut {NoStop}%
\bibitem [{\citenamefont {Ullmann}(1976)}]{ullmann_picture_1976}%
  \BibitemOpen
  \bibfield  {author} {\bibinfo {author} {\bibfnamefont {J.~R.}\ \bibnamefont
  {Ullmann}},\ }\bibfield  {title} {\bibinfo {title} {Picture analysis in
  character recognition},\ }in\ \href {https://doi.org/10.1007/3540075798_24}
  {\emph {\bibinfo {booktitle} {Digital {Picture} {Analysis}}}},\ \bibinfo
  {editor} {edited by\ \bibinfo {editor} {\bibfnamefont {A.}~\bibnamefont
  {Rosenfeld}}}\ (\bibinfo  {publisher} {Springer},\ \bibinfo {address}
  {Berlin, Heidelberg},\ \bibinfo {year} {1976})\ pp.\ \bibinfo {pages}
  {295--343}\BibitemShut {NoStop}%
\bibitem [{\citenamefont {Huang}(1974)}]{huang_noise_1974}%
  \BibitemOpen
  \bibfield  {author} {\bibinfo {author} {\bibfnamefont {G.~C.}\ \bibnamefont
  {Huang}},\ }\bibfield  {title} {\bibinfo {title} {Noise {Reduction} by
  {Adaptive} {Threshold} in {Digital} {Signal} {Processing}},\ }in\ \href
  {https://doi.org/10.1109/ISEMC.1974.7567118} {\emph {\bibinfo {booktitle}
  {Proc. {ISEMC} '74}}}\ (\bibinfo {address} {San Francisco, CA, USA},\
  \bibinfo {year} {1974})\ pp.\ \bibinfo {pages} {1--7}\BibitemShut {NoStop}%
\bibitem [{\citenamefont {Daubechies}(1992)}]{daubechies_ten_1992}%
  \BibitemOpen
  \bibfield  {author} {\bibinfo {author} {\bibfnamefont {I.}~\bibnamefont
  {Daubechies}},\ }\href
  {http://epubs.siam.org/doi/book/10.1137/1.9781611970104} {\emph {\bibinfo
  {title} {Ten {Lectures} on {Wavelets}}}},\ {CBMS}-{NSF} {Regional}
  {Conference} {Series} in {Applied} {Mathematics}\ (\bibinfo  {publisher}
  {Society for Industrial and Applied Mathematics},\ \bibinfo {year}
  {1992})\BibitemShut {NoStop}%
\bibitem [{\citenamefont {Donoho}\ and\ \citenamefont
  {Johnstone}(1994)}]{donoho_ideal_1994}%
  \BibitemOpen
  \bibfield  {author} {\bibinfo {author} {\bibfnamefont {D.~L.}\ \bibnamefont
  {Donoho}}\ and\ \bibinfo {author} {\bibfnamefont {I.~M.}\ \bibnamefont
  {Johnstone}},\ }\bibfield  {title} {\bibinfo {title} {Ideal spatial
  adaptation by wavelet shrinkage},\ }\href
  {https://doi.org/10.1093/biomet/81.3.425} {\bibfield  {journal} {\bibinfo
  {journal} {Biometrika}\ }\textbf {\bibinfo {volume} {81}},\ \bibinfo {pages}
  {425} (\bibinfo {year} {1994})}\BibitemShut {NoStop}%
\bibitem [{\citenamefont {Donoho}\ \emph {et~al.}(1995)\citenamefont {Donoho},
  \citenamefont {Johnstone}, \citenamefont {Kerkyacharian},\ and\ \citenamefont
  {Picard}}]{donoho_wavelet_1995}%
  \BibitemOpen
  \bibfield  {author} {\bibinfo {author} {\bibfnamefont {D.~L.}\ \bibnamefont
  {Donoho}}, \bibinfo {author} {\bibfnamefont {I.~M.}\ \bibnamefont
  {Johnstone}}, \bibinfo {author} {\bibfnamefont {G.}~\bibnamefont
  {Kerkyacharian}},\ and\ \bibinfo {author} {\bibfnamefont {D.}~\bibnamefont
  {Picard}},\ }\bibfield  {title} {\bibinfo {title} {Wavelet {Shrinkage}:
  {Asymptopia}?},\ }\href {https://doi.org/10.1111/j.2517-6161.1995.tb02032.x}
  {\bibfield  {journal} {\bibinfo  {journal} {J. Royal Stat. Soc. B}\ }\textbf
  {\bibinfo {volume} {57}},\ \bibinfo {pages} {301} (\bibinfo {year}
  {1995})}\BibitemShut {NoStop}%
\bibitem [{\citenamefont {Rudin}\ \emph {et~al.}(1992)\citenamefont {Rudin},
  \citenamefont {Osher},\ and\ \citenamefont {Fatemi}}]{rudin_nonlinear_1992}%
  \BibitemOpen
  \bibfield  {author} {\bibinfo {author} {\bibfnamefont {L.~I.}\ \bibnamefont
  {Rudin}}, \bibinfo {author} {\bibfnamefont {S.}~\bibnamefont {Osher}},\ and\
  \bibinfo {author} {\bibfnamefont {E.}~\bibnamefont {Fatemi}},\ }\bibfield
  {title} {\bibinfo {title} {Nonlinear total variation based noise removal
  algorithms},\ }\href {https://doi.org/10.1016/0167-2789(92)90242-F}
  {\bibfield  {journal} {\bibinfo  {journal} {Physica D}\ }\textbf {\bibinfo
  {volume} {60}},\ \bibinfo {pages} {259} (\bibinfo {year} {1992})}\BibitemShut
  {NoStop}%
\bibitem [{\citenamefont {Strong}\ and\ \citenamefont
  {Chan}(2003)}]{strong_edge-preserving_2003}%
  \BibitemOpen
  \bibfield  {author} {\bibinfo {author} {\bibfnamefont {D.}~\bibnamefont
  {Strong}}\ and\ \bibinfo {author} {\bibfnamefont {T.}~\bibnamefont {Chan}},\
  }\bibfield  {title} {\bibinfo {title} {Edge-preserving and scale-dependent
  properties of total variation regularization},\ }\href
  {https://doi.org/10.1088/0266-5611/19/6/059} {\bibfield  {journal} {\bibinfo
  {journal} {Inverse Probl.}\ }\textbf {\bibinfo {volume} {19}},\ \bibinfo
  {pages} {S165} (\bibinfo {year} {2003})}\BibitemShut {NoStop}%
\bibitem [{\citenamefont {Caselles}\ \emph {et~al.}(2007)\citenamefont
  {Caselles}, \citenamefont {Chambolle},\ and\ \citenamefont
  {Novaga}}]{caselles_discontinuity_2007}%
  \BibitemOpen
  \bibfield  {author} {\bibinfo {author} {\bibfnamefont {V.}~\bibnamefont
  {Caselles}}, \bibinfo {author} {\bibfnamefont {A.}~\bibnamefont
  {Chambolle}},\ and\ \bibinfo {author} {\bibfnamefont {M.}~\bibnamefont
  {Novaga}},\ }\bibfield  {title} {\bibinfo {title} {The {Discontinuity} {Set}
  of {Solutions} of the {TV} {Denoising} {Problem} and {Some} {Extensions}},\
  }\href {https://doi.org/10.1137/070683003} {\bibfield  {journal} {\bibinfo
  {journal} {Multiscale Model. Simul.}\ }\textbf {\bibinfo {volume} {6}},\
  \bibinfo {pages} {879} (\bibinfo {year} {2007})}\BibitemShut {NoStop}%
\bibitem [{\citenamefont {Cleveland}(1979)}]{cleveland_robust_1979}%
  \BibitemOpen
  \bibfield  {author} {\bibinfo {author} {\bibfnamefont {W.~S.}\ \bibnamefont
  {Cleveland}},\ }\bibfield  {title} {\bibinfo {title} {Robust {Locally}
  {Weighted} {Regression} and {Smoothing} {Scatterplots}},\ }\href
  {https://doi.org/10.1080/01621459.1979.10481038} {\bibfield  {journal}
  {\bibinfo  {journal} {J. Am. Stat. Assoc.}\ }\textbf {\bibinfo {volume}
  {74}},\ \bibinfo {pages} {829} (\bibinfo {year} {1979})}\BibitemShut
  {NoStop}%
\bibitem [{\citenamefont {Savitzky}\ and\ \citenamefont
  {Golay}(1964)}]{savitzky_smoothing_1964}%
  \BibitemOpen
  \bibfield  {author} {\bibinfo {author} {\bibfnamefont {A.}~\bibnamefont
  {Savitzky}}\ and\ \bibinfo {author} {\bibfnamefont {M.~J.~E.}\ \bibnamefont
  {Golay}},\ }\bibfield  {title} {\bibinfo {title} {Smoothing and
  {Differentiation} of {Data} by {Simplified} {Least} {Squares} {Procedures}},\
  }\href {https://doi.org/10.1021/ac60214a047} {\bibfield  {journal} {\bibinfo
  {journal} {Anal. Chem.}\ }\textbf {\bibinfo {volume} {36}},\ \bibinfo {pages}
  {1627} (\bibinfo {year} {1964})}\BibitemShut {NoStop}%
\bibitem [{\citenamefont {Wiener}(1949)}]{wiener_extrapolation_1949}%
  \BibitemOpen
  \bibfield  {author} {\bibinfo {author} {\bibfnamefont {N.}~\bibnamefont
  {Wiener}},\ }\href@noop {} {\emph {\bibinfo {title} {Extrapolation,
  interpolation, and smoothing of stationary time series: with engineering
  applications.}}}\ (\bibinfo  {publisher} {MIT Press},\ \bibinfo {address}
  {Cambridge},\ \bibinfo {year} {1949})\BibitemShut {NoStop}%
\bibitem [{\citenamefont {Kalman}(1960)}]{kalman_new_1960}%
  \BibitemOpen
  \bibfield  {author} {\bibinfo {author} {\bibfnamefont {R.~E.}\ \bibnamefont
  {Kalman}},\ }\bibfield  {title} {\bibinfo {title} {A {New} {Approach} to
  {Linear} {Filtering} and {Prediction} {Problems}},\ }\href
  {https://doi.org/10.1115/1.3662552} {\bibfield  {journal} {\bibinfo
  {journal} {J. Basic Eng.}\ }\textbf {\bibinfo {volume} {82}},\ \bibinfo
  {pages} {35} (\bibinfo {year} {1960})}\BibitemShut {NoStop}%
\bibitem [{\citenamefont {Stewart}(1993)}]{stewart_early_1993}%
  \BibitemOpen
  \bibfield  {author} {\bibinfo {author} {\bibfnamefont {G.}~\bibnamefont
  {Stewart}},\ }\bibfield  {title} {\bibinfo {title} {On the {Early} {History}
  of the {Singular} {Value} {Decomposition}},\ }\href
  {https://doi.org/10.1137/1035134} {\bibfield  {journal} {\bibinfo  {journal}
  {SIAM Rev.}\ }\textbf {\bibinfo {volume} {35}},\ \bibinfo {pages} {551}
  (\bibinfo {year} {1993})}\BibitemShut {NoStop}%
\bibitem [{\citenamefont {Pearson}(1901)}]{pearson_liii._1901}%
  \BibitemOpen
  \bibfield  {author} {\bibinfo {author} {\bibfnamefont {K.~F.}\ \bibnamefont
  {Pearson}},\ }\bibfield  {title} {\bibinfo {title} {{LIII}. {On} lines and
  planes of closest fit to systems of points in space},\ }\href
  {https://doi.org/10.1080/14786440109462720} {\bibfield  {journal} {\bibinfo
  {journal} {Philos. Mag.}\ }\textbf {\bibinfo {volume} {2}},\ \bibinfo {pages}
  {559} (\bibinfo {year} {1901})}\BibitemShut {NoStop}%
\bibitem [{\citenamefont {Andrews}\ and\ \citenamefont
  {Patterson}(1976)}]{andrews_singular_1976}%
  \BibitemOpen
  \bibfield  {author} {\bibinfo {author} {\bibfnamefont {H.}~\bibnamefont
  {Andrews}}\ and\ \bibinfo {author} {\bibfnamefont {C.}~\bibnamefont
  {Patterson}},\ }\bibfield  {title} {\bibinfo {title} {Singular {Value}
  {Decomposition} ({SVD}) {Image} {Coding}},\ }\href
  {https://doi.org/10.1109/TCOM.1976.1093309} {\bibfield  {journal} {\bibinfo
  {journal} {IEEE Trans. Commun.}\ }\textbf {\bibinfo {volume} {24}},\ \bibinfo
  {pages} {425} (\bibinfo {year} {1976})}\BibitemShut {NoStop}%
\bibitem [{\citenamefont {Chang}\ \emph {et~al.}(2005)\citenamefont {Chang},
  \citenamefont {Tsai},\ and\ \citenamefont {Lin}}]{chang_svd-based_2005}%
  \BibitemOpen
  \bibfield  {author} {\bibinfo {author} {\bibfnamefont {C.-C.}\ \bibnamefont
  {Chang}}, \bibinfo {author} {\bibfnamefont {P.}~\bibnamefont {Tsai}},\ and\
  \bibinfo {author} {\bibfnamefont {C.-C.}\ \bibnamefont {Lin}},\ }\bibfield
  {title} {\bibinfo {title} {{SVD}-based digital image watermarking scheme},\
  }\href {https://doi.org/10.1016/j.patrec.2005.01.004} {\bibfield  {journal}
  {\bibinfo  {journal} {Pattern Recognit. Lett.}\ }\textbf {\bibinfo {volume}
  {26}},\ \bibinfo {pages} {1577} (\bibinfo {year} {2005})}\BibitemShut
  {NoStop}%
\bibitem [{\citenamefont {Guo}\ \emph {et~al.}(2016)\citenamefont {Guo},
  \citenamefont {Zhang}, \citenamefont {Zhang},\ and\ \citenamefont
  {Liu}}]{guo_efficient_2016}%
  \BibitemOpen
  \bibfield  {author} {\bibinfo {author} {\bibfnamefont {Q.}~\bibnamefont
  {Guo}}, \bibinfo {author} {\bibfnamefont {C.}~\bibnamefont {Zhang}}, \bibinfo
  {author} {\bibfnamefont {Y.}~\bibnamefont {Zhang}},\ and\ \bibinfo {author}
  {\bibfnamefont {H.}~\bibnamefont {Liu}},\ }\bibfield  {title} {\bibinfo
  {title} {An {Efficient} {SVD}-{Based} {Method} for {Image} {Denoising}},\
  }\href {https://doi.org/10.1109/TCSVT.2015.2416631} {\bibfield  {journal}
  {\bibinfo  {journal} {IEEE Trans. Circuits Syst. Video Technol.}\ }\textbf
  {\bibinfo {volume} {26}},\ \bibinfo {pages} {868} (\bibinfo {year}
  {2016})}\BibitemShut {NoStop}%
\bibitem [{\citenamefont {Furnival}\ \emph {et~al.}(2017)\citenamefont
  {Furnival}, \citenamefont {Leary},\ and\ \citenamefont
  {Midgley}}]{furnival_denoising_2017}%
  \BibitemOpen
  \bibfield  {author} {\bibinfo {author} {\bibfnamefont {T.}~\bibnamefont
  {Furnival}}, \bibinfo {author} {\bibfnamefont {R.~K.}\ \bibnamefont
  {Leary}},\ and\ \bibinfo {author} {\bibfnamefont {P.~A.}\ \bibnamefont
  {Midgley}},\ }\bibfield  {title} {\bibinfo {title} {Denoising time-resolved
  microscopy image sequences with singular value thresholding},\ }\href
  {https://doi.org/10.1016/j.ultramic.2016.05.005} {\bibfield  {journal}
  {\bibinfo  {journal} {Ultramicroscopy}\ }\textbf {\bibinfo {volume} {178}},\
  \bibinfo {pages} {112} (\bibinfo {year} {2017})}\BibitemShut {NoStop}%
\bibitem [{\citenamefont {Hermus}\ \emph {et~al.}(1999)\citenamefont {Hermus},
  \citenamefont {Dologlou}, \citenamefont {Wambacq},\ and\ \citenamefont
  {Compernolle}}]{hermus_fully_1999}%
  \BibitemOpen
  \bibfield  {author} {\bibinfo {author} {\bibfnamefont {K.}~\bibnamefont
  {Hermus}}, \bibinfo {author} {\bibfnamefont {I.}~\bibnamefont {Dologlou}},
  \bibinfo {author} {\bibfnamefont {P.}~\bibnamefont {Wambacq}},\ and\ \bibinfo
  {author} {\bibfnamefont {D.~V.}\ \bibnamefont {Compernolle}},\ }\bibfield
  {title} {\bibinfo {title} {Fully adaptive {SVD}-based noise removal for
  robust speech recognition},\ }in\ \href
  {https://www.isca-speech.org/archive/eurospeech_1999/e99_1951.html} {\emph
  {\bibinfo {booktitle} {Proc. {EUROSPEECH} '99}}}\ (\bibinfo {address}
  {Budapest, Hungary},\ \bibinfo {year} {1999})\ pp.\ \bibinfo {pages}
  {1951--1954}\BibitemShut {NoStop}%
\bibitem [{\citenamefont {Al-Zaben}\ and\ \citenamefont
  {Al-Smadi}(2006)}]{al-zaben_extraction_2006}%
  \BibitemOpen
  \bibfield  {author} {\bibinfo {author} {\bibfnamefont {A.}~\bibnamefont
  {Al-Zaben}}\ and\ \bibinfo {author} {\bibfnamefont {A.}~\bibnamefont
  {Al-Smadi}},\ }\bibfield  {title} {\bibinfo {title} {Extraction of foetal
  {ECG} by combination of singular value decomposition and neuro-fuzzy
  inference system},\ }\href {https://doi.org/10.1088/0031-9155/51/1/010}
  {\bibfield  {journal} {\bibinfo  {journal} {Phys. Med. Biol.}\ }\textbf
  {\bibinfo {volume} {51}},\ \bibinfo {pages} {137} (\bibinfo {year}
  {2006})}\BibitemShut {NoStop}%
\bibitem [{\citenamefont {Chiron}\ \emph {et~al.}(2014)\citenamefont {Chiron},
  \citenamefont {van Agthoven}, \citenamefont {Kieffer}, \citenamefont
  {Rolando},\ and\ \citenamefont {Delsuc}}]{chiron_efficient_2014}%
  \BibitemOpen
  \bibfield  {author} {\bibinfo {author} {\bibfnamefont {L.}~\bibnamefont
  {Chiron}}, \bibinfo {author} {\bibfnamefont {M.~A.}\ \bibnamefont {van
  Agthoven}}, \bibinfo {author} {\bibfnamefont {B.}~\bibnamefont {Kieffer}},
  \bibinfo {author} {\bibfnamefont {C.}~\bibnamefont {Rolando}},\ and\ \bibinfo
  {author} {\bibfnamefont {M.-A.}\ \bibnamefont {Delsuc}},\ }\bibfield  {title}
  {\bibinfo {title} {Efficient denoising algorithms for large experimental
  datasets and their applications in {Fourier} transform ion cyclotron
  resonance mass spectrometry},\ }\href
  {https://doi.org/10.1073/pnas.1306700111} {\bibfield  {journal} {\bibinfo
  {journal} {Proc. Natl. Acad. Sci.}\ }\textbf {\bibinfo {volume} {111}},\
  \bibinfo {pages} {1385} (\bibinfo {year} {2014})}\BibitemShut {NoStop}%
\bibitem [{\citenamefont {Cong}\ \emph {et~al.}(2013)\citenamefont {Cong},
  \citenamefont {Chen}, \citenamefont {Dong},\ and\ \citenamefont
  {Zhao}}]{cong_short-time_2013}%
  \BibitemOpen
  \bibfield  {author} {\bibinfo {author} {\bibfnamefont {F.}~\bibnamefont
  {Cong}}, \bibinfo {author} {\bibfnamefont {J.}~\bibnamefont {Chen}}, \bibinfo
  {author} {\bibfnamefont {G.}~\bibnamefont {Dong}},\ and\ \bibinfo {author}
  {\bibfnamefont {F.}~\bibnamefont {Zhao}},\ }\bibfield  {title} {\bibinfo
  {title} {Short-time matrix series based singular value decomposition for
  rolling bearing fault diagnosis},\ }\href
  {https://doi.org/10.1016/j.ymssp.2012.06.005} {\bibfield  {journal} {\bibinfo
   {journal} {Mech. Syst. Signal Process.}\ }\textbf {\bibinfo {volume} {34}},\
  \bibinfo {pages} {218} (\bibinfo {year} {2013})}\BibitemShut {NoStop}%
\bibitem [{\citenamefont {Zhao}\ and\ \citenamefont
  {Jia}(2017)}]{zhao_novel_2017}%
  \BibitemOpen
  \bibfield  {author} {\bibinfo {author} {\bibfnamefont {M.}~\bibnamefont
  {Zhao}}\ and\ \bibinfo {author} {\bibfnamefont {X.}~\bibnamefont {Jia}},\
  }\bibfield  {title} {\bibinfo {title} {A novel strategy for signal denoising
  using reweighted {SVD} and its applications to weak fault feature enhancement
  of rotating machinery},\ }\href {https://doi.org/10.1016/j.ymssp.2017.02.036}
  {\bibfield  {journal} {\bibinfo  {journal} {Mech. Syst. Signal Process.}\
  }\textbf {\bibinfo {volume} {94}},\ \bibinfo {pages} {129} (\bibinfo {year}
  {2017})}\BibitemShut {NoStop}%
\bibitem [{\citenamefont {Zhao}\ and\ \citenamefont
  {Ye}(2011)}]{zhao_selection_2011}%
  \BibitemOpen
  \bibfield  {author} {\bibinfo {author} {\bibfnamefont {X.}~\bibnamefont
  {Zhao}}\ and\ \bibinfo {author} {\bibfnamefont {B.}~\bibnamefont {Ye}},\
  }\bibfield  {title} {\bibinfo {title} {Selection of effective singular values
  using difference spectrum and its application to fault diagnosis of
  headstock},\ }\href {https://doi.org/10.1016/j.ymssp.2011.01.003} {\bibfield
  {journal} {\bibinfo  {journal} {Mech. Syst. Signal Process.}\ }\textbf
  {\bibinfo {volume} {25}},\ \bibinfo {pages} {1617} (\bibinfo {year}
  {2011})}\BibitemShut {NoStop}%
\bibitem [{\citenamefont {Jiang}\ \emph {et~al.}(2015)\citenamefont {Jiang},
  \citenamefont {Chen}, \citenamefont {Dong}, \citenamefont {Liu},\ and\
  \citenamefont {Chen}}]{jiang_study_2015}%
  \BibitemOpen
  \bibfield  {author} {\bibinfo {author} {\bibfnamefont {H.}~\bibnamefont
  {Jiang}}, \bibinfo {author} {\bibfnamefont {J.}~\bibnamefont {Chen}},
  \bibinfo {author} {\bibfnamefont {G.}~\bibnamefont {Dong}}, \bibinfo {author}
  {\bibfnamefont {T.}~\bibnamefont {Liu}},\ and\ \bibinfo {author}
  {\bibfnamefont {G.}~\bibnamefont {Chen}},\ }\bibfield  {title} {\bibinfo
  {title} {Study on {Hankel} matrix-based {SVD} and its application in rolling
  element bearing fault diagnosis},\ }\href
  {https://doi.org/10.1016/j.ymssp.2014.07.019} {\bibfield  {journal} {\bibinfo
   {journal} {Mech. Syst. Signal Process.}\ }\textbf {\bibinfo {volume}
  {52-53}},\ \bibinfo {pages} {338} (\bibinfo {year} {2015})}\BibitemShut
  {NoStop}%
\bibitem [{\citenamefont {Schanze}(2018)}]{schanze_compression_2018}%
  \BibitemOpen
  \bibfield  {author} {\bibinfo {author} {\bibfnamefont {T.}~\bibnamefont
  {Schanze}},\ }\bibfield  {title} {\bibinfo {title} {Compression and {Noise}
  {Reduction} of {Biomedical} {Signals} by {Singular} {Value}
  {Decomposition}},\ }\href {https://doi.org/10.1016/j.ifacol.2018.03.062}
  {\bibfield  {journal} {\bibinfo  {journal} {IFAC-PapersOnLine}\ }\textbf
  {\bibinfo {volume} {51}},\ \bibinfo {pages} {361} (\bibinfo {year}
  {2018})}\BibitemShut {NoStop}%
\bibitem [{\citenamefont {Hassanpour}(2008)}]{hassanpour_timefrequency_2008}%
  \BibitemOpen
  \bibfield  {author} {\bibinfo {author} {\bibfnamefont {H.}~\bibnamefont
  {Hassanpour}},\ }\bibfield  {title} {\bibinfo {title} {A time-frequency
  approach for noise reduction},\ }\href
  {https://doi.org/10.1016/j.dsp.2007.09.014} {\bibfield  {journal} {\bibinfo
  {journal} {Digit. Signal Process.}\ }\textbf {\bibinfo {volume} {18}},\
  \bibinfo {pages} {728} (\bibinfo {year} {2008})}\BibitemShut {NoStop}%
\bibitem [{\citenamefont {Gong}\ \emph {et~al.}(2017)\citenamefont {Gong},
  \citenamefont {Li},\ and\ \citenamefont {Zhao}}]{gong_improved_2017}%
  \BibitemOpen
  \bibfield  {author} {\bibinfo {author} {\bibfnamefont {W.}~\bibnamefont
  {Gong}}, \bibinfo {author} {\bibfnamefont {H.}~\bibnamefont {Li}},\ and\
  \bibinfo {author} {\bibfnamefont {D.}~\bibnamefont {Zhao}},\ }\bibfield
  {title} {\bibinfo {title} {An {Improved} {Denoising} {Model} {Based} on the
  {Analysis} {K}-{SVD} {Algorithm}},\ }\href
  {https://doi.org/10.1007/s00034-017-0496-7} {\bibfield  {journal} {\bibinfo
  {journal} {Circuits Syst. Signal Process.}\ }\textbf {\bibinfo {volume}
  {36}},\ \bibinfo {pages} {4006} (\bibinfo {year} {2017})}\BibitemShut
  {NoStop}%
\bibitem [{\citenamefont {Shin}\ \emph {et~al.}(1999)\citenamefont {Shin},
  \citenamefont {Hammond},\ and\ \citenamefont {White}}]{shin_iterative_1999}%
  \BibitemOpen
  \bibfield  {author} {\bibinfo {author} {\bibfnamefont {K.}~\bibnamefont
  {Shin}}, \bibinfo {author} {\bibfnamefont {J.}~\bibnamefont {Hammond}},\ and\
  \bibinfo {author} {\bibfnamefont {P.}~\bibnamefont {White}},\ }\bibfield
  {title} {\bibinfo {title} {Iterative {SVD} method for noise reduction of
  low-dimensional chaotic time series},\ }\href
  {https://doi.org/10.1006/mssp.1998.9999} {\bibfield  {journal} {\bibinfo
  {journal} {Mech. Syst. Signal Process.}\ }\textbf {\bibinfo {volume} {13}},\
  \bibinfo {pages} {115} (\bibinfo {year} {1999})}\BibitemShut {NoStop}%
\bibitem [{\citenamefont {Eckart}\ and\ \citenamefont
  {Young}(1936)}]{eckart_approximation_1936}%
  \BibitemOpen
  \bibfield  {author} {\bibinfo {author} {\bibfnamefont {C.}~\bibnamefont
  {Eckart}}\ and\ \bibinfo {author} {\bibfnamefont {G.}~\bibnamefont {Young}},\
  }\bibfield  {title} {\bibinfo {title} {The approximation of one matrix by
  another of lower rank},\ }\href {https://doi.org/10.1007/BF02288367}
  {\bibfield  {journal} {\bibinfo  {journal} {Psychometrika}\ }\textbf
  {\bibinfo {volume} {1}},\ \bibinfo {pages} {211} (\bibinfo {year}
  {1936})}\BibitemShut {NoStop}%
\bibitem [{\citenamefont {Yang}\ and\ \citenamefont
  {Tse}(2003)}]{yang_development_2003}%
  \BibitemOpen
  \bibfield  {author} {\bibinfo {author} {\bibfnamefont {W.-X.}\ \bibnamefont
  {Yang}}\ and\ \bibinfo {author} {\bibfnamefont {P.~W.}\ \bibnamefont {Tse}},\
  }\bibfield  {title} {\bibinfo {title} {Development of an advanced noise
  reduction method for vibration analysis based on singular value
  decomposition},\ }\href {https://doi.org/10.1016/S0963-8695(03)00044-6}
  {\bibfield  {journal} {\bibinfo  {journal} {NDT E Int.}\ }\textbf {\bibinfo
  {volume} {36}},\ \bibinfo {pages} {419} (\bibinfo {year} {2003})}\BibitemShut
  {NoStop}%
\bibitem [{\citenamefont {Candès}\ and\ \citenamefont
  {Recht}(2009)}]{candes_exact_2009}%
  \BibitemOpen
  \bibfield  {author} {\bibinfo {author} {\bibfnamefont {E.~J.}\ \bibnamefont
  {Candès}}\ and\ \bibinfo {author} {\bibfnamefont {B.}~\bibnamefont
  {Recht}},\ }\bibfield  {title} {\bibinfo {title} {Exact {Matrix} {Completion}
  via {Convex} {Optimization}},\ }\href
  {https://doi.org/10.1007/s10208-009-9045-5} {\bibfield  {journal} {\bibinfo
  {journal} {Found. Comput. Math.}\ }\textbf {\bibinfo {volume} {9}},\ \bibinfo
  {pages} {717} (\bibinfo {year} {2009})}\BibitemShut {NoStop}%
\bibitem [{\citenamefont {Karner}\ \emph {et~al.}(2003)\citenamefont {Karner},
  \citenamefont {Schneid},\ and\ \citenamefont
  {Ueberhuber}}]{karner_spectral_2003}%
  \BibitemOpen
  \bibfield  {author} {\bibinfo {author} {\bibfnamefont {H.}~\bibnamefont
  {Karner}}, \bibinfo {author} {\bibfnamefont {J.}~\bibnamefont {Schneid}},\
  and\ \bibinfo {author} {\bibfnamefont {C.~W.}\ \bibnamefont {Ueberhuber}},\
  }\bibfield  {title} {\bibinfo {title} {Spectral decomposition of real
  circulant matrices},\ }\href {https://doi.org/10.1016/S0024-3795(02)00664-X}
  {\bibfield  {journal} {\bibinfo  {journal} {Linear Algebra Appl.}\ }\textbf
  {\bibinfo {volume} {367}},\ \bibinfo {pages} {301} (\bibinfo {year}
  {2003})}\BibitemShut {NoStop}%
\bibitem [{\citenamefont {Gibbs}(1898)}]{gibbs_fouriers_1898}%
  \BibitemOpen
  \bibfield  {author} {\bibinfo {author} {\bibfnamefont {J.~W.}\ \bibnamefont
  {Gibbs}},\ }\bibfield  {title} {\bibinfo {title} {Fourier's {Series}},\
  }\href {https://doi.org/10.1038/059200b0} {\bibfield  {journal} {\bibinfo
  {journal} {Nature}\ }\textbf {\bibinfo {volume} {59}},\ \bibinfo {pages}
  {200} (\bibinfo {year} {1898})}\BibitemShut {NoStop}%
\bibitem [{\citenamefont {Duffing}(1918)}]{duffing_erzwungene_1918}%
  \BibitemOpen
  \bibfield  {author} {\bibinfo {author} {\bibfnamefont {G.}~\bibnamefont
  {Duffing}},\ }\href@noop {} {\emph {\bibinfo {title} {Erzwungene
  {Schwingungen} bei veränderlicher {Eigenfrequenz} und ihre technische
  {Bedeutung}}}},\ \bibinfo {series} {Sammlung {Vieweg}}\ No.\ \bibinfo
  {number} {41--42}\ (\bibinfo  {publisher} {F. Vieweg \& sohn},\ \bibinfo
  {address} {Braunschweig},\ \bibinfo {year} {1918})\BibitemShut {NoStop}%
\bibitem [{\citenamefont {Franzke}\ \emph {et~al.}(2008)\citenamefont
  {Franzke}, \citenamefont {Geissel},\ and\ \citenamefont
  {Münzenberg}}]{franzke_mass_2008}%
  \BibitemOpen
  \bibfield  {author} {\bibinfo {author} {\bibfnamefont {B.}~\bibnamefont
  {Franzke}}, \bibinfo {author} {\bibfnamefont {H.}~\bibnamefont {Geissel}},\
  and\ \bibinfo {author} {\bibfnamefont {G.}~\bibnamefont {Münzenberg}},\
  }\bibfield  {title} {\bibinfo {title} {Mass and lifetime measurements of
  exotic nuclei in storage rings},\ }\href {https://doi.org/10.1002/mas.20173}
  {\bibfield  {journal} {\bibinfo  {journal} {Mass Spectrom. Rev.}\ }\textbf
  {\bibinfo {volume} {27}},\ \bibinfo {pages} {428} (\bibinfo {year}
  {2008})}\BibitemShut {NoStop}%
\bibitem [{\citenamefont {Hausmann}\ \emph {et~al.}(2000)\citenamefont
  {Hausmann} \emph {et~al.}}]{hausmann_first_2000}%
  \BibitemOpen
  \bibfield  {author} {\bibinfo {author} {\bibfnamefont {M.}~\bibnamefont
  {Hausmann}} \emph {et~al.},\ }\bibfield  {title} {\bibinfo {title} {First
  isochronous mass spectrometry at the experimental storage ring {ESR}},\
  }\href {https://doi.org/10.1016/S0168-9002(99)01192-4} {\bibfield  {journal}
  {\bibinfo  {journal} {Nucl. Instrum. Meth. A}\ }\textbf {\bibinfo {volume}
  {446}},\ \bibinfo {pages} {569} (\bibinfo {year} {2000})}\BibitemShut
  {NoStop}%
\bibitem [{\citenamefont {Xu}\ \emph {et~al.}(2013)\citenamefont {Xu},
  \citenamefont {Zhang},\ and\ \citenamefont {Litvinov}}]{xu_accurate_2013}%
  \BibitemOpen
  \bibfield  {author} {\bibinfo {author} {\bibfnamefont {H.~S.}\ \bibnamefont
  {Xu}}, \bibinfo {author} {\bibfnamefont {Y.~H.}\ \bibnamefont {Zhang}},\ and\
  \bibinfo {author} {\bibfnamefont {Y.~A.}\ \bibnamefont {Litvinov}},\
  }\bibfield  {title} {\bibinfo {title} {Accurate mass measurements of exotic
  nuclei with the {CSRe} in {Lanzhou}},\ }\href
  {https://doi.org/10.1016/j.ijms.2013.04.029} {\bibfield  {journal} {\bibinfo
  {journal} {Int. J. Mass Spectrom.}\ }\textbf {\bibinfo {volume} {349--350}},\
  \bibinfo {pages} {162} (\bibinfo {year} {2013})}\BibitemShut {NoStop}%
\bibitem [{\citenamefont {Mei}\ \emph {et~al.}(2010)\citenamefont {Mei} \emph
  {et~al.}}]{mei_high_2010}%
  \BibitemOpen
  \bibfield  {author} {\bibinfo {author} {\bibfnamefont {B.}~\bibnamefont
  {Mei}} \emph {et~al.},\ }\bibfield  {title} {\bibinfo {title} {A high
  performance time-of-flight detector applied to isochronous mass measurement
  at {CSRe}},\ }\href {https://doi.org/10.1016/j.nima.2010.09.001} {\bibfield
  {journal} {\bibinfo  {journal} {Nucl. Instrum. Meth. A}\ }\textbf {\bibinfo
  {volume} {624}},\ \bibinfo {pages} {109} (\bibinfo {year}
  {2010})}\BibitemShut {NoStop}%
\bibitem [{\citenamefont {Tu}\ \emph {et~al.}(2011)\citenamefont {Tu} \emph
  {et~al.}}]{tu_direct_2011}%
  \BibitemOpen
  \bibfield  {author} {\bibinfo {author} {\bibfnamefont {X.~L.}\ \bibnamefont
  {Tu}} \emph {et~al.},\ }\bibfield  {title} {\bibinfo {title} {Direct mass
  measurements of short-lived {$A=2Z-1$} nuclides {$^{{63}}$Ge}, {$^{{65}}$As},
  {$^{{67}}$Se}, and {$^{{71}}$Kr} and their impact on nucleosynthesis in the
  {$rp$} process},\ }\href {https://doi.org/10.1103/PhysRevLett.106.112501}
  {\bibfield  {journal} {\bibinfo  {journal} {Phys. Rev. Lett.}\ }\textbf
  {\bibinfo {volume} {106}},\ \bibinfo {pages} {112501} (\bibinfo {year}
  {2011})}\BibitemShut {NoStop}%
\bibitem [{\citenamefont {Zhang}\ \emph {et~al.}(2012)\citenamefont {Zhang}
  \emph {et~al.}}]{zhang_mass_2012}%
  \BibitemOpen
  \bibfield  {author} {\bibinfo {author} {\bibfnamefont {Y.~H.}\ \bibnamefont
  {Zhang}} \emph {et~al.},\ }\bibfield  {title} {\bibinfo {title} {Mass
  measurements of the neutron-deficient {$^{{41}}$Ti}, {$^{{45}}$Cr},
  {$^{{49}}$Fe}, and {$^{{53}}$Ni} nuclides: {First} test of the isobaric
  multiplet mass equation in {$fp$}-shell nuclei},\ }\href
  {https://doi.org/10.1103/PhysRevLett.109.102501} {\bibfield  {journal}
  {\bibinfo  {journal} {Phys. Rev. Lett.}\ }\textbf {\bibinfo {volume} {109}},\
  \bibinfo {pages} {102501} (\bibinfo {year} {2012})}\BibitemShut {NoStop}%
\bibitem [{\citenamefont {Xu}\ \emph {et~al.}(2016)\citenamefont {Xu} \emph
  {et~al.}}]{xu_identification_2016}%
  \BibitemOpen
  \bibfield  {author} {\bibinfo {author} {\bibfnamefont {X.}~\bibnamefont {Xu}}
  \emph {et~al.},\ }\bibfield  {title} {\bibinfo {title} {Identification of the
  {Lowest} {$T=2,J^\pi=0^+$} {Isobaric} {Analog} {State} in {$^{{52}}$Co} and
  {Its} {Impact} on the {Understanding} of {$\beta$-Decay} {Properties} of
  {$^{{52}}$Ni}},\ }\href {https://doi.org/10.1103/PhysRevLett.117.182503}
  {\bibfield  {journal} {\bibinfo  {journal} {Phys. Rev. Lett.}\ }\textbf
  {\bibinfo {volume} {117}},\ \bibinfo {pages} {182503} (\bibinfo {year}
  {2016})}\BibitemShut {NoStop}%
\bibitem [{\citenamefont {Breiman}(1995)}]{breiman_better_1995}%
  \BibitemOpen
  \bibfield  {author} {\bibinfo {author} {\bibfnamefont {L.}~\bibnamefont
  {Breiman}},\ }\bibfield  {title} {\bibinfo {title} {Better {Subset}
  {Regression} {Using} the {Nonnegative} {Garrote}},\ }\href
  {https://doi.org/10.1080/00401706.1995.10484371} {\bibfield  {journal}
  {\bibinfo  {journal} {Technometrics}\ }\textbf {\bibinfo {volume} {37}},\
  \bibinfo {pages} {373} (\bibinfo {year} {1995})}\BibitemShut {NoStop}%
\bibitem [{\citenamefont {Gao}(1998)}]{gao_wavelet_1998}%
  \BibitemOpen
  \bibfield  {author} {\bibinfo {author} {\bibfnamefont {H.-Y.}\ \bibnamefont
  {Gao}},\ }\bibfield  {title} {\bibinfo {title} {Wavelet {Shrinkage}
  {Denoising} {Using} the {Non}-{Negative} {Garrote}},\ }\href
  {https://doi.org/10.1080/10618600.1998.10474789} {\bibfield  {journal}
  {\bibinfo  {journal} {J. Comput. Graph. Stat.}\ }\textbf {\bibinfo {volume}
  {7}},\ \bibinfo {pages} {469} (\bibinfo {year} {1998})}\BibitemShut {NoStop}%
\bibitem [{\citenamefont {Lee}\ \emph {et~al.}(2006)\citenamefont {Lee},
  \citenamefont {Gommers}, \citenamefont {Wasilewski}, \citenamefont
  {Wohlfahrt}, \citenamefont {O'Leary}, \citenamefont {Nahrstaedt},\ and\
  \citenamefont {Contributors}}]{lee_pywavelets_2006}%
  \BibitemOpen
  \bibfield  {author} {\bibinfo {author} {\bibfnamefont {G.}~\bibnamefont
  {Lee}}, \bibinfo {author} {\bibfnamefont {R.}~\bibnamefont {Gommers}},
  \bibinfo {author} {\bibfnamefont {F.}~\bibnamefont {Wasilewski}}, \bibinfo
  {author} {\bibfnamefont {K.}~\bibnamefont {Wohlfahrt}}, \bibinfo {author}
  {\bibfnamefont {A.}~\bibnamefont {O'Leary}}, \bibinfo {author} {\bibfnamefont
  {H.}~\bibnamefont {Nahrstaedt}},\ and\ \bibinfo {author} {\bibnamefont
  {Contributors}},\ }\href {https://github.com/PyWavelets/pywt} {\bibinfo
  {title} {{PyWavelets}---{Wavelet} {Transforms} in {Python}}},\ \bibinfo
  {howpublished} {GitHub} (\bibinfo {year} {2006})\BibitemShut {NoStop}%
\bibitem [{\citenamefont {Condat}(2013)}]{condat_direct_2013}%
  \BibitemOpen
  \bibfield  {author} {\bibinfo {author} {\bibfnamefont {L.}~\bibnamefont
  {Condat}},\ }\bibfield  {title} {\bibinfo {title} {A {Direct} {Algorithm} for
  1-{D} {Total} {Variation} {Denoising}},\ }\href
  {https://doi.org/10.1109/LSP.2013.2278339} {\bibfield  {journal} {\bibinfo
  {journal} {IEEE Signal Process. Lett.}\ }\textbf {\bibinfo {volume} {20}},\
  \bibinfo {pages} {1054} (\bibinfo {year} {2013})}\BibitemShut {NoStop}%
\bibitem [{\citenamefont {Condat}(2017)}]{condat_total_2017}%
  \BibitemOpen
  \bibfield  {author} {\bibinfo {author} {\bibfnamefont {L.}~\bibnamefont
  {Condat}},\ }\href
  {http://www.gipsa-lab.grenoble-inp.fr/~laurent.condat/software.html}
  {\bibinfo {title} {Total variation denoising of {1-D} signals---{Version}
  2.0}} (\bibinfo {year} {2017})\BibitemShut {NoStop}%
\bibitem [{\citenamefont {Zhang}\ \emph {et~al.}(2014)\citenamefont {Zhang}
  \emph {et~al.}}]{zhang_timing_2014}%
  \BibitemOpen
  \bibfield  {author} {\bibinfo {author} {\bibfnamefont {W.}~\bibnamefont
  {Zhang}} \emph {et~al.},\ }\bibfield  {title} {\bibinfo {title} {A timing
  detector with pulsed high-voltage power supply for mass measurements at
  {CSRe}},\ }\href {https://doi.org/10.1016/j.nima.2014.04.031} {\bibfield
  {journal} {\bibinfo  {journal} {Nucl. Instrum. Meth. A}\ }\textbf {\bibinfo
  {volume} {755}},\ \bibinfo {pages} {38} (\bibinfo {year} {2014})}\BibitemShut
  {NoStop}%
\bibitem [{\citenamefont {Litvinov}\ \emph {et~al.}(2005)\citenamefont
  {Litvinov} \emph {et~al.}}]{litvinov_mass_2005}%
  \BibitemOpen
  \bibfield  {author} {\bibinfo {author} {\bibfnamefont {Y.~A.}\ \bibnamefont
  {Litvinov}} \emph {et~al.},\ }\bibfield  {title} {\bibinfo {title} {Mass
  measurement of cooled neutron-deficient bismuth projectile fragments with
  time-resolved {Schottky} mass spectrometry at the {FRS}-{ESR} facility},\
  }\href {https://doi.org/10.1016/j.nuclphysa.2005.03.015} {\bibfield
  {journal} {\bibinfo  {journal} {Nucl. Phys. A}\ }\textbf {\bibinfo {volume}
  {756}},\ \bibinfo {pages} {3} (\bibinfo {year} {2005})}\BibitemShut {NoStop}%
\bibitem [{\citenamefont {Nolden}\ \emph {et~al.}(2011)\citenamefont {Nolden}
  \emph {et~al.}}]{nolden_fast_2011}%
  \BibitemOpen
  \bibfield  {author} {\bibinfo {author} {\bibfnamefont {F.}~\bibnamefont
  {Nolden}} \emph {et~al.},\ }\bibfield  {title} {\bibinfo {title} {A fast and
  sensitive resonant {Schottky} pick-up for heavy ion storage rings},\ }\href
  {https://doi.org/10.1016/j.nima.2011.06.058} {\bibfield  {journal} {\bibinfo
  {journal} {Nucl. Instrum. Meth. A}\ }\textbf {\bibinfo {volume} {659}},\
  \bibinfo {pages} {69} (\bibinfo {year} {2011})}\BibitemShut {NoStop}%
\bibitem [{\citenamefont
  {Chattopadhyay}(1985)}]{chattopadhyay_fundamental_1985}%
  \BibitemOpen
  \bibfield  {author} {\bibinfo {author} {\bibfnamefont {S.}~\bibnamefont
  {Chattopadhyay}},\ }\bibfield  {title} {\bibinfo {title} {Some fundamental
  aspects of fluctuations and coherence in charged‐particle beams in storage
  rings},\ }\href {https://doi.org/10.1063/1.35175} {\bibfield  {journal}
  {\bibinfo  {journal} {AIP Conf. Proc.}\ }\textbf {\bibinfo {volume} {127}},\
  \bibinfo {pages} {467} (\bibinfo {year} {1985})}\BibitemShut {NoStop}%
\bibitem [{\citenamefont {Trageser}\ \emph {et~al.}(2015)\citenamefont
  {Trageser} \emph {et~al.}}]{trageser_new_2015}%
  \BibitemOpen
  \bibfield  {author} {\bibinfo {author} {\bibfnamefont {C.}~\bibnamefont
  {Trageser}} \emph {et~al.},\ }\bibfield  {title} {\bibinfo {title} {A new
  data acquisition system for {Schottky} signals in atomic physics experiments
  at {GSI}'s and {FAIR}'s storage rings},\ }\href
  {https://doi.org/10.1088/0031-8949/2015/T166/014062} {\bibfield  {journal}
  {\bibinfo  {journal} {Phys. Scr.}\ }\textbf {\bibinfo {volume} {2015}},\
  \bibinfo {pages} {014062} (\bibinfo {year} {2015})}\BibitemShut {NoStop}%
\bibitem [{\citenamefont {Litvinov}\ and\ \citenamefont
  {Bosch}(2011)}]{litvinov_beta_2011}%
  \BibitemOpen
  \bibfield  {author} {\bibinfo {author} {\bibfnamefont {Y.~A.}\ \bibnamefont
  {Litvinov}}\ and\ \bibinfo {author} {\bibfnamefont {F.}~\bibnamefont
  {Bosch}},\ }\bibfield  {title} {\bibinfo {title} {Beta decay of highly
  charged ions},\ }\href {https://doi.org/10.1088/0034-4885/74/1/016301}
  {\bibfield  {journal} {\bibinfo  {journal} {Rep. Prog. Phys.}\ }\textbf
  {\bibinfo {volume} {74}},\ \bibinfo {pages} {016301} (\bibinfo {year}
  {2011})}\BibitemShut {NoStop}%
\bibitem [{\citenamefont {Tu}\ \emph {et~al.}(2018)\citenamefont {Tu} \emph
  {et~al.}}]{tu_first_2018}%
  \BibitemOpen
  \bibfield  {author} {\bibinfo {author} {\bibfnamefont {X.~L.}\ \bibnamefont
  {Tu}} \emph {et~al.},\ }\bibfield  {title} {\bibinfo {title} {First
  application of combined isochronous and {Schottky} mass spectrometry:
  {Half}-lives of fully ionized {$^{{49}}$Cr$^{{24+}}$} and
  {$^{{53}}$Fe$^{{26+}}$} atoms},\ }\href
  {https://doi.org/10.1103/PhysRevC.97.014321} {\bibfield  {journal} {\bibinfo
  {journal} {Phys. Rev. C}\ }\textbf {\bibinfo {volume} {97}},\ \bibinfo
  {pages} {014321} (\bibinfo {year} {2018})}\BibitemShut {NoStop}%
\bibitem [{\citenamefont {Chen}\ \emph {et~al.}(2015)\citenamefont {Chen} \emph
  {et~al.}}]{chen_accuracy_2015}%
  \BibitemOpen
  \bibfield  {author} {\bibinfo {author} {\bibfnamefont {X.}~\bibnamefont
  {Chen}} \emph {et~al.},\ }\bibfield  {title} {\bibinfo {title} {Accuracy
  improvement in the isochronous mass measurement using a cavity doublet},\
  }\href {https://doi.org/10.1007/s10751-015-1183-3} {\bibfield  {journal}
  {\bibinfo  {journal} {Hyperfine Interact.}\ }\textbf {\bibinfo {volume}
  {235}},\ \bibinfo {pages} {51} (\bibinfo {year} {2015})}\BibitemShut
  {NoStop}%
\bibitem [{\citenamefont {Chen}(2015)}]{chen_non-interceptive_2015}%
  \BibitemOpen
  \bibfield  {author} {\bibinfo {author} {\bibfnamefont {X.}~\bibnamefont
  {Chen}},\ }\emph {\bibinfo {title} {Non-interceptive position detection for
  short-lived radioactive nuclei in heavy-ion storage rings}},\ \href
  {http://www.ub.uni-heidelberg.de/archiv/19851} {Ph.D. thesis},\ \bibinfo
  {school} {Ruprecht-Karls-Universit\"at Heidelberg} (\bibinfo {year}
  {2015})\BibitemShut {NoStop}%
\bibitem [{\citenamefont {Schoukens}\ and\ \citenamefont
  {Renneboog}(1986)}]{schoukens_modeling_1986}%
  \BibitemOpen
  \bibfield  {author} {\bibinfo {author} {\bibfnamefont {J.}~\bibnamefont
  {Schoukens}}\ and\ \bibinfo {author} {\bibfnamefont {J.}~\bibnamefont
  {Renneboog}},\ }\bibfield  {title} {\bibinfo {title} {Modeling the noise
  influence on the {Fourier} coefficients after a discrete {Fourier}
  transform},\ }\href {https://doi.org/10.1109/TIM.1986.6499210} {\bibfield
  {journal} {\bibinfo  {journal} {IEEE Trans. Instrum. Meas.}\ }\textbf
  {\bibinfo {volume} {IM-35}},\ \bibinfo {pages} {278} (\bibinfo {year}
  {1986})}\BibitemShut {NoStop}%
\bibitem [{\citenamefont {Goodman}(1975)}]{goodman_statistical_1975}%
  \BibitemOpen
  \bibfield  {author} {\bibinfo {author} {\bibfnamefont {J.~W.}\ \bibnamefont
  {Goodman}},\ }\bibfield  {title} {\bibinfo {title} {Statistical {Properties}
  of {Laser} {Speckle} {Patterns}},\ }in\ \href
  {https://doi.org/10.1007/978-3-662-43205-1_2} {\emph {\bibinfo {booktitle}
  {Laser {Speckle} and {Related} {Phenomena}}}},\ \bibinfo {editor} {edited by\
  \bibinfo {editor} {\bibfnamefont {J.~C.}\ \bibnamefont {Dainty}}}\ (\bibinfo
  {publisher} {Springer},\ \bibinfo {address} {Berlin, Heidelberg},\ \bibinfo
  {year} {1975})\ pp.\ \bibinfo {pages} {9--75}\BibitemShut {NoStop}%
\bibitem [{\citenamefont {Arsenault}\ and\ \citenamefont
  {Denis}(1983)}]{arsenault_image_1983}%
  \BibitemOpen
  \bibfield  {author} {\bibinfo {author} {\bibfnamefont {H.~H.}\ \bibnamefont
  {Arsenault}}\ and\ \bibinfo {author} {\bibfnamefont {M.}~\bibnamefont
  {Denis}},\ }\bibfield  {title} {\bibinfo {title} {Image processing in
  signal-dependent noise},\ }\href {https://doi.org/10.1139/p83-042} {\bibfield
   {journal} {\bibinfo  {journal} {Can. J. Phys.}\ }\textbf {\bibinfo {volume}
  {61}},\ \bibinfo {pages} {309} (\bibinfo {year} {1983})}\BibitemShut
  {NoStop}%
\bibitem [{\citenamefont {Litvinov}\ \emph {et~al.}(2004)\citenamefont
  {Litvinov} \emph {et~al.}}]{litvinov_precision_2004}%
  \BibitemOpen
  \bibfield  {author} {\bibinfo {author} {\bibfnamefont {Y.~A.}\ \bibnamefont
  {Litvinov}} \emph {et~al.},\ }\bibfield  {title} {\bibinfo {title} {Precision
  experiments with time-resolved {Schottky} mass spectrometry},\ }\href
  {https://doi.org/10.1016/j.nuclphysa.2004.01.089} {\bibfield  {journal}
  {\bibinfo  {journal} {Nucl. Phys. A}\ }\textbf {\bibinfo {volume} {734}},\
  \bibinfo {pages} {473} (\bibinfo {year} {2004})}\BibitemShut {NoStop}%
\bibitem [{\citenamefont {Kienle}\ \emph {et~al.}(2013)\citenamefont {Kienle}
  \emph {et~al.}}]{kienle_high-resolution_2013}%
  \BibitemOpen
  \bibfield  {author} {\bibinfo {author} {\bibfnamefont {P.}~\bibnamefont
  {Kienle}} \emph {et~al.},\ }\bibfield  {title} {\bibinfo {title}
  {High-resolution measurement of the time-modulated orbital electron capture
  and of the {$\beta^+$} decay of hydrogen-like {$^{{142}}$Pm$^{{60+}}$}
  ions},\ }\href {https://doi.org/10.1016/j.physletb.2013.09.033} {\bibfield
  {journal} {\bibinfo  {journal} {Phys. Lett. B}\ }\textbf {\bibinfo {volume}
  {726}},\ \bibinfo {pages} {638} (\bibinfo {year} {2013})}\BibitemShut
  {NoStop}%
\bibitem [{\citenamefont {Walker}\ \emph {et~al.}(2013)\citenamefont {Walker},
  \citenamefont {Litvinov},\ and\ \citenamefont {Geissel}}]{walker_ilima_2013}%
  \BibitemOpen
  \bibfield  {author} {\bibinfo {author} {\bibfnamefont {P.~M.}\ \bibnamefont
  {Walker}}, \bibinfo {author} {\bibfnamefont {Y.~A.}\ \bibnamefont
  {Litvinov}},\ and\ \bibinfo {author} {\bibfnamefont {H.}~\bibnamefont
  {Geissel}},\ }\bibfield  {title} {\bibinfo {title} {The {ILIMA} project at
  {FAIR}},\ }\href {https://doi.org/10.1016/j.ijms.2013.04.007} {\bibfield
  {journal} {\bibinfo  {journal} {Int. J. Mass Spectrom.}\ }\textbf {\bibinfo
  {volume} {349--350}},\ \bibinfo {pages} {247} (\bibinfo {year}
  {2013})}\BibitemShut {NoStop}%
\bibitem [{\citenamefont {Chen}(2017)}]{chen_band-limited_2017}%
  \BibitemOpen
  \bibfield  {author} {\bibinfo {author} {\bibfnamefont {X.}~\bibnamefont
  {Chen}},\ }\bibfield  {title} {\bibinfo {title} {Band-limited peak-finding
  method for a noisy frequency spectrum},\ }in\ \href@noop {} {\emph {\bibinfo
  {booktitle} {Proc. {STORI} '17}}}\ (\bibinfo {address} {Kanazawa, Japan},\
  \bibinfo {year} {2017})\BibitemShut {NoStop}%
\bibitem [{\citenamefont {Litvinov}\ \emph {et~al.}(2007)\citenamefont
  {Litvinov} \emph {et~al.}}]{litvinov_measurement_2007}%
  \BibitemOpen
  \bibfield  {author} {\bibinfo {author} {\bibfnamefont {Y.~A.}\ \bibnamefont
  {Litvinov}} \emph {et~al.},\ }\bibfield  {title} {\bibinfo {title}
  {Measurement of the {$\beta^+$} and orbital electron-capture decay rates in
  fully ionized, hydrogenlike, and heliumlike {$^{{140}}$Pr} ions},\ }\href
  {https://doi.org/10.1103/PhysRevLett.99.262501} {\bibfield  {journal}
  {\bibinfo  {journal} {Phys. Rev. Lett.}\ }\textbf {\bibinfo {volume} {99}},\
  \bibinfo {pages} {262501} (\bibinfo {year} {2007})}\BibitemShut {NoStop}%
\bibitem [{\citenamefont {Bosch}\ \emph {et~al.}(1996)\citenamefont {Bosch}
  \emph {et~al.}}]{bosch_observation_1996}%
  \BibitemOpen
  \bibfield  {author} {\bibinfo {author} {\bibfnamefont {F.}~\bibnamefont
  {Bosch}} \emph {et~al.},\ }\bibfield  {title} {\bibinfo {title} {Observation
  of bound-state {$\beta^-$} decay of fully ionized {$^{{187}}$Re}:
  {$^{{187}}$Re}-{$^{{187}}$Os} cosmochronometry},\ }\href
  {https://doi.org/10.1103/PhysRevLett.77.5190} {\bibfield  {journal} {\bibinfo
   {journal} {Phys. Rev. Lett.}\ }\textbf {\bibinfo {volume} {77}},\ \bibinfo
  {pages} {5190} (\bibinfo {year} {1996})}\BibitemShut {NoStop}%
\bibitem [{\citenamefont {Chen}(2019)}]{chen_generic_2019}%
  \BibitemOpen
  \bibfield  {author} {\bibinfo {author} {\bibfnamefont {X.}~\bibnamefont
  {Chen}},\ }\href {https://doi.org/10.5281/zenodo.2603558} {\bibinfo {title}
  {A generic denoising method for {1D} spectra based on singular value
  decomposition}},\ \bibinfo {howpublished} {Zenodo} (\bibinfo {year}
  {2019})\BibitemShut {NoStop}%
\end{thebibliography}
\end{document}